\documentclass[12pt]{article}

\usepackage{newtxtext,newtxmath}

\usepackage{graphicx}

\usepackage[letterpaper,margin=1in]{geometry}

\renewenvironment{abstract}
	{\quotation}
	{\endquotation}

\date{}

\makeatletter
\renewcommand{\fnum@figure}{\textbf{Figure \thefigure}}
\renewcommand{\fnum@table}{\textbf{Table \thetable}}
\makeatother

\usepackage{scicite}

\usepackage{url}

\def\scititle{An Encoder-Decoder Foundation Chemical Language Model for Generative Polymer Design
}
\title{\bfseries \boldmath \scititle}

\author{
	Harikrishna Sahu$^{1}$,
        Wei Xiong$^{1}$,
        Anagha Savit$^{1}$,
        Shivank S Shukla$^{1}$,
	Rampi Ramprasad$^{1\ast}$\and
	\small$^{1}$School of Materials Science and Engineering, Georgia Institute of Technology, Atlanta, GA 30332, USA\and
	\small$^\ast$Corresponding author. Email: rampi.ramprasad@mse.gatech.edu\and
}

\begin{document} 

\maketitle

\begin{abstract} \bfseries \boldmath
Traditional machine learning has advanced polymer discovery, yet direct generation of chemically valid and synthesizable polymers without exhaustive enumeration remains a challenge. Here we present \textsc{polyT5}, an encoder–decoder chemical language model based on the T5 architecture, trained to understand and generate polymer structures. \textsc{polyT5} enables both property prediction and the targeted generation of polymers conditioned on desired property values. We demonstrate its utility for dielectric polymer design, seeking candidates with dielectric constant $>$3, bandgap $>$4 eV, and glass transition temperature $>$400 K, alongside melt-processability and solubility requirements. From over 20,000 generated promising candidates, one was experimentally synthesized and validated, showing strong agreement with predictions. To further enhance usability, we integrated \textsc{polyT5} within an agentic AI framework that couples it with a general-purpose LLM, allowing natural language interaction for property prediction and generative design. Together, these advances establish a versatile and accessible framework for accelerated polymer discovery.
\end{abstract}

\noindent

\newpage
Materials discovery is undergoing a transformation with the rise of generative models—particularly large language models (LLMs)—for tasks such as data extraction from scientific literature, property prediction, synthesis planning, and molecular generation.\cite{zha25:14, pei25:119419, ram25:2514, pyz25:61, van25:20743, ans23:8736} A particular class of generative schemes, referred to as ``foundation" models, are trained on massive general-purpose language datasets, requiring large vocabularies and architectures with billions of parameters. While powerful, they demand substantial computational resources and lack the domain-specific knowledge essential for materials science. Even when fine-tuned for materials applications, these models can hallucinate, producing syntactically valid but chemically implausible or infeasible outputs.\cite{zak24:313} Training on curated datasets encoded with domain-appropriate tokens makes it likely that these models will outperform general-purpose counterparts while using only a fraction of the parameters, reducing computational cost and enhancing reliability, interpretability, and adaptability for downstream materials discovery tasks.\cite{zha25:202500085, liu23:10688, che25:04267}

Advances in generative models for materials discovery have been preceded  by the development of the so-called  state-of-the-art (SOTA) models. These predictive models have traditionally relied on conventional machine learning (ML) techniques, which typically require carefully engineered numerical fingerprints to represent chemical structure. More recently, general-purpose LLMs fine-tuned for specific prediction tasks\cite{ram25:2514, ani25:100184} offer advantages over SOTA models by bypassing complex feature engineering through natural language prompts, handling missing features more effectively, and adapting rapidly to new problems. In our recent work, we applied this fine-tuning strategy to polymers across multiple property domains, including thermal properties\cite{gup25:02129} and solubility\cite{aga25:2017}, demonstrating its potential to streamline predictive modeling in materials informatics.


While accurate property prediction is crucial, the complementary challenge in materials discovery is design, i.e., generating candidate structures that meet desired property or performance criteria. In high-throughput screening powered by ML, candidates\textemdash first generated by users based on heuristics\textemdash are rapidly evaluated to identify promising materials. However, these approaches still explore only a minute fraction of the vast chemical space and remain constrained by user-imposed biases and limitations inherent in the enumeration process.\cite{che20:163, rey15:722, sah19:17480} Generative design methods address this limitation by automatically producing structures conditioned on target properties, enabling more efficient navigation of chemical space and accelerating the identification of globally optimal materials. A variety of deep generative modeling approaches have been explored for molecular design, including variational autoencoders (VAEs), generative adversarial networks (GANs), graph neural networks (GNNs), and more recently, transformer-based architectures.\cite{par24:2355, ans23:8736} Dollar et al.\cite{dol21:8362} showed that incorporating self-attention layers into generative VAE models enables them to learn complex molecular grammar, with individual attention heads capturing distinct high-level relationships between atomic and structural groups, thereby improving the handling of longer and more complex SMILES (Simplified Molecular Input Line Entry System) strings. He et al.\cite{he21:26} demonstrated the effectiveness of transformer-based models for translating molecules into property-optimized counterparts using SMILES representations. Chemformer,\cite{irw22:015022} a transformer-based model leveraging SMILES representations, demonstrated that self-supervised pre-training and transfer learning can enable efficient fine-tuning across diverse sequence-to-sequence and discriminative cheminformatics tasks, achieving state-of-the-art performance in synthesis prediction and molecular optimization.

For generative molecular design, the choice of string representation for chemical structures is critically important. While SMILES\cite{dav88:31, dav89:97} has long been the de facto standard for encoding molecules/polymers, its susceptibility to syntactic and semantic errors often leads to invalid structures when used in combination with generative algorithms.\cite{sat21:7607, bat20:10489, ski24:437} This limitation can hinder the efficiency and reliability of design workflows. In contrast, SELFIES (SELF-referencIng Embedded Strings)\cite{kre20:045024} provides a 100\% robust alternative that guarantees syntactically and semantically valid chemical structures, irrespective of how the string is modified or generated. This robustness enables the creation of a valid and continuous latent chemical space, making it particularly well-suited for deep generative models. Owing to its robustness, SELFIES is increasingly being adopted in generative molecular design.\cite{kre22:100588, xu24:353, pio23:25868, alb24:1} 

Since SELFIES was originally developed for small molecules, its direct application to polymers was limited. To overcome this, in our recent work we introduced a pseudo-SELFIES notation specifically tailored for polymers, where the termini (*) were substituted with Astatine atoms (At). This molecule-like representation enabled the further training of the SELFIES-TED model\cite{ibm25:selfiested}\textemdash originally designed for molecules but adapted here for polymers\textemdash leading to the development of polyBART\cite{ana25:04233}. The model was then used to design candidate polymers by perturbing the latent space of known structures with targeted properties. However, the approach inherently constrained the generated candidates to remain close to the starting polymer within the learned chemical space.

Despite growing interest in applying deep generative models to materials discovery, critical gaps remain in the domain of complex organic materials\textemdash especially polymers. Given their structural diversity and chemical complexity, there is a strong need for a foundation LLM trained specifically on polymers, capable of capturing underlying structural relationships that can be transferred to downstream tasks. From a methodological perspective, many current language models rely on decoder-only architectures, which process sequences unidirectionally and therefore fail to capture the bidirectional chemical dependencies critical for understanding polymer structures. Moreover, common token-level masking strategies are insufficient for learning high-level structural patterns, as they do not capture multi-token dependencies and long-range contextual relationships essential for effective polymer modeling.

The Text-to-Text Transfer Transformer (T5), an encoder–decoder model developed by Raffel et al.,\cite{col23:10683} addresses these key limitations through bidirectional encoder attention and a span-masking strategy with sentinel tokens. T5-based models have shown strong potential in chemical domains: MolT5\cite{edw22:2204} translates between molecular representations and natural language, enabling molecular captioning and design from textual descriptions; C5T5\cite{dan21:10307} optimizes properties from IUPAC names for drug discovery; and other variants facilitate forward and retrosynthetic planning.\cite{lu22:1376} While T5 has achieved initial success for organic materials, its general-purpose nature limits performance, highlighting the need for a polymer-specific base model attuned to the structural and chemical nuances of polymers.

Here, we present \textsc{polyT5}, a domain-adapted encoder–decoder language model based on the T5 architecture, trained on over 100 million chemically diverse polymer structures in SELFIES representation of known and hypothetical candidates, spanning 20 functional groups, 8 heteroatoms, with broad representation of aromatic rings and aliphatic chains. Leveraging bidirectional encoder attention, \textsc{polyT5} captures long-range structural relationships, while its span-masking strategy more effectively learns missing and next-token patterns than conventional token-level masking. 

\begin{figure} 
	\centering
	\includegraphics[width=0.9\textwidth]{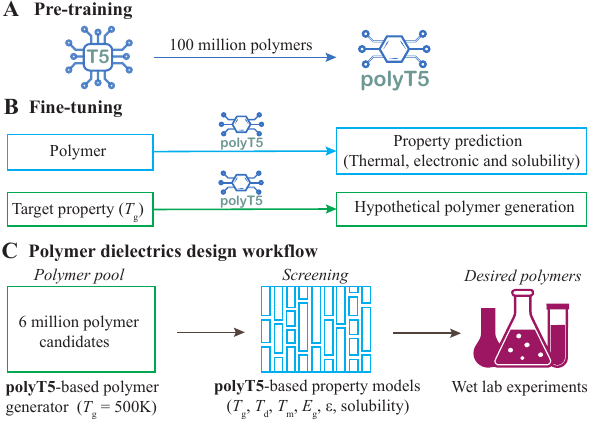} 

        \caption{\textbf{Schematic workflow illustrating the large language model (LLM)-based framework for organic material design targeting dielectric applications.} 
        (\textbf{A}) \textsc{polyT5}, a T5-based language model, pre-trained on a corpus of 100 million polymer structures to capture underlying chemical and structural patterns.
        (\textbf{B}) Fine-tuning \textsc{polyT5} for domain-specific tasks, including: (i) property prediction for thermal, electrical, and solubility-related properties and (ii) generative design — generating hypothetical polymer candidates based on a target property (e.g., glass transition temperature, $T_{\rm g}$)).
        (\textbf{C}) Application of the fine-tuned model for dielectric polymer design by generating hypothetical polymers targeting a $T_{\rm g}$ of 500~K, followed by screening to identify candidates satisfying dielectric, thermal stability, and solubility criteria, thus enabling a data-driven materials discovery pipeline.
        }
	\label{fig1}
\end{figure}

To demonstrate its practical utility, \textsc{polyT5} was applied to dielectric polymer design, an important class of materials for advanced electronic and energy applications. The base model was fine-tuned for two key tasks: (i) hypothetical polymer generation conditioned on target glass transition temperature ($T_{\rm g}$) and (ii) prediction of thermal, electronic, and solubility properties. This framework generated over 6 million candidates, which were screened to identify polymers with high dielectric constant, wide bandgap, elevated $T_{\rm g}$, and favorable melt-processability and solubility. A schematic of this LLM-driven pipeline for dielectric polymer design is shown in Figure~\ref{fig1}. A top-performing candidate was synthesized and experimentally validated, with measured properties matching predictions, establishing an end-to-end, informatics-driven workflow for accelerated generative discovery of next-generation functional polymers. To further broaden accessibility, we also demonstrate integration of \textsc{polyT5} within an agentic AI framework that combines it with a general-purpose LLM, enabling natural language interaction for property prediction and generative design.


\subsection*{\textsc{PolyT5}: Foundation model for polymers}
To develop a foundation model for polymers capable of understanding intrinsic structure–property relationships, the first step is to curate a sufficiently large and chemically diverse training dataset. To this end, a large representative dataset of homopolymer chemical structures was constructed, including 12,473 experimentally synthesized polymers collected from the literature and over 100 million hypothetically generated polymers. To ensure chemical validity and synthetic relevance, the hypothetical polymers were generated by reacting commercially available molecules using well-established reactions such as polycondensation,\cite{ gur24:6107, kim23:318, ohn23:5539, yue24:2465} click chemistry,\cite{kl01:2004, gen21:963} and ring-opening metathesis polymerization (ROMP)\cite{odi04:544, gur24:6107}. These polymers encompass a broad range of functional groups, heteroatoms, and degrees of backbone saturation, as summarized in Table \ref{tabS1}.

\begin{figure} 
	\centering
	\includegraphics[width=1.0\textwidth]{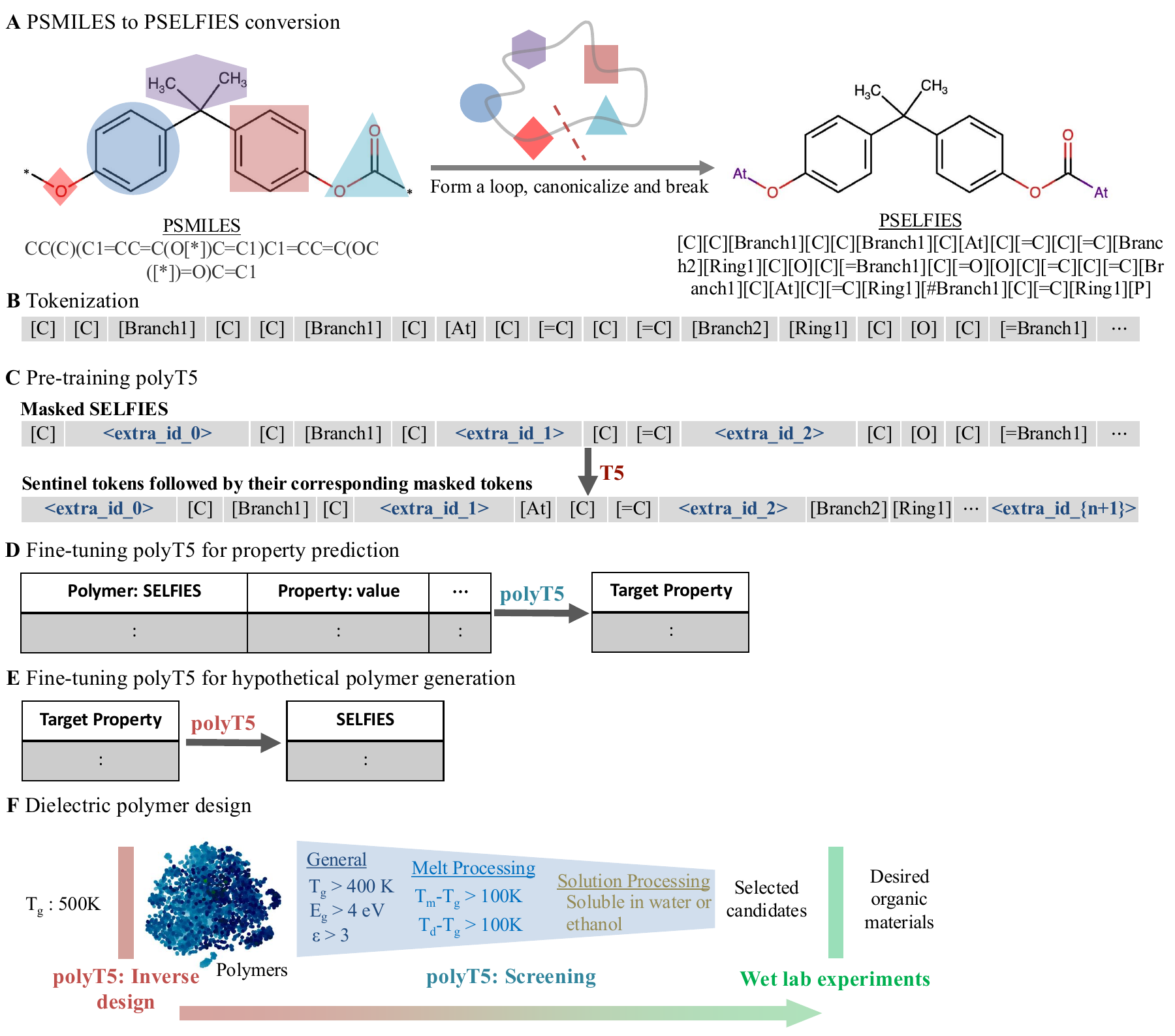} 

	\caption{\textbf{\textsc{polyT5}: Key steps in the design of polymers for dielectric applications.}
    (\textbf{A}) Conversion of polymer SMILES (PSMILES) to polymer SELFIES (PSELFIES) representations.
    (\textbf{B}) Tokenization of PSELFIES for language model input.
    (\textbf{C}) Implementation of a masking strategy for pre-training the base \textsc{polyT5} model.
    (\textbf{D}) Fine-tuning of the base model for property prediction tasks, including thermal, electronic, and solubility properties.
    (\textbf{E}) Fine-tuning of the base model for conditional hypothetical polymer generation.
    (\textbf{F}) Generation and screening of hypothetical polymers targeting desired dielectric properties through integrated property prediction models.}
	\label{fig2}
\end{figure}

Each polymer was initially represented using polymer-specific SMILES (PSMILES) notation, where two \textbf{[*]}s denote the terminal ends of the polymer chain. However, since \textbf{[*]} is conventionally used to denote dummy atoms and is not supported in SELFIES—an inherently robust representation well suited for generative modeling of molecules—we developed a custom conversion strategy to overcome this limitation, as illustrated in Figure~\ref{fig2}A.\cite{ana25:04233} During this process, the two polymer ends were first joined by removing the \textbf{[*]} tokens to form a cyclic structure, followed by canonicalization to mitigate any initial sequence bias. Subsequently, a single bond within the backbone was strategically cleaved, and Astatine (\textbf{At}) atoms were attached at the cleavage points. \textbf{At} was selected because it is rarely encountered in polymer structures and was entirely absent from our training dataset, minimizing the risk of unintended bias. The resulting pseudo-polymer SMILES\textemdash molecular representations with \textbf{At} atoms marking the polymer termini\textemdash were subsequently converted to SELFIES\cite{kre20:045024}, referred to as PSELFIES. This curated dataset served as the foundation for pre-training the base \textsc{polyT5} language model. A custom tokenizer vocabulary was defined to support this representation, comprising 458 tokens including SELFIES tokens (see Figure~\ref{fig2}B), special markers, and additional tokens for property-conditional generation, all integrated as predefined tokens to ensure compatibility with the SentencePiece framework.\cite{methods} 

To explore the effect of model capacity on downstream performance, we pre-trained three variants of the T5 architecture with progressively increasing sizes, referred to as Small, Medium, and Large.\cite{methods} These models differ in the number of transformer layers, embedding dimensions, and attention heads, with total parameters ranging from approximately 1.4 million to 59 million (see Table~\ref{tabS0}). Pre-training was carried out using a span-masking strategy in which segments of SELFIES strings were replaced with sentinel tokens (Figure~\ref{fig2}C). Across all models, the maximum sequence length was fixed at 200 positions, which is sufficient to cover 99.91\% of the SELFIES tokens present in the training dataset, as shown in Figure~\ref{token_count}. This setup enabled a systematic assessment of the trade-off between computational efficiency and representational power for polymer informatics applications.

Figure~\ref{train_loss} shows the training loss per batch across two epochs for all three \textsc{polyT5} model variants. The loss decreases sharply within the first 0.5 epoch and then plateaus, likely due to the limited token vocabulary (199 unique tokens) used to represent over 100 million PSELFIES, as well as the inherent structural similarity among polymers, despite the underlying chemical diversity present in the training dataset. As expected, the \textsc{polyT5}-small model exhibits higher loss, while the medium and large variants show marginal differences.

\subsection*{Downstream Task 1: Property Prediction}
The base \textsc{polyT5} models were fine-tuned for a variety of property prediction tasks (Figure~\ref{fig2}D), covering both regression and classification cases, across thermal, electronic, and solubility properties. The training datasets used for fine-tuning, comprising experimentally measured or previously computed polymer properties, were compiled from earlier studies. The corresponding property distributions are shown in Figure~\ref{prop_train}.\cite{ana25:04233, kue23:4099} The glass transition temperature ($T_{\rm g}$) dataset includes 5,130 polymers, with values ranging from 80 K to 873 K and a median around 394 K. The distribution is slightly right-skewed, with a higher concentration of polymers between 325 K and 488 K, as shown in Figure~\ref{prop_train}(A). For decomposition temperature ($T_d$), the dataset contains 4,204 entries, spanning from 291 K to 1,167 K. The melting temperature ($T_m$) dataset, comprising 2,151 polymers, displays a slightly narrower range, from 226 K to 860 K, with most values falling between 400 K and 550 K. The electronic bandgap ($E_g$) dataset, computed using density functional theory (DFT), consists of 4,113 polymers, covering values from as low as 0.07 eV to nearly 9.84 eV. The majority of polymers exhibit $E_g$ values clustered between 3.4 eV and 5.6 eV, with a median value of 4.58 eV (Figure~\ref{prop_train}(D)). The dielectric constant ($\varepsilon$) dataset comprises 1,569 polymers, with values ranging from 1.68 to 10.40. The distribution is moderately skewed toward higher values, with a median of 3.09 and most polymers falling between 2.55 and 3.97. The dataset includes both computational and experimental values: 382 dielectric constants were calculated using DFT, while the remaining 1,187 values were experimentally measured at nine different frequencies, ranging from 60 Hz to $10^{15}$ Hz. The distribution of $\varepsilon$ values at each measurement frequency, along with the DFT-calculated and total combined distributions, is provided in Figure~\ref{prop_train}(E).  For polymer–solvent solubility, the dataset contains 19,245 soluble and 9,970 insoluble cases, covering 6,246 unique polymers tested across 58 different solvents.

\begin{figure} 
	\centering
	\includegraphics[width=0.9\textwidth]{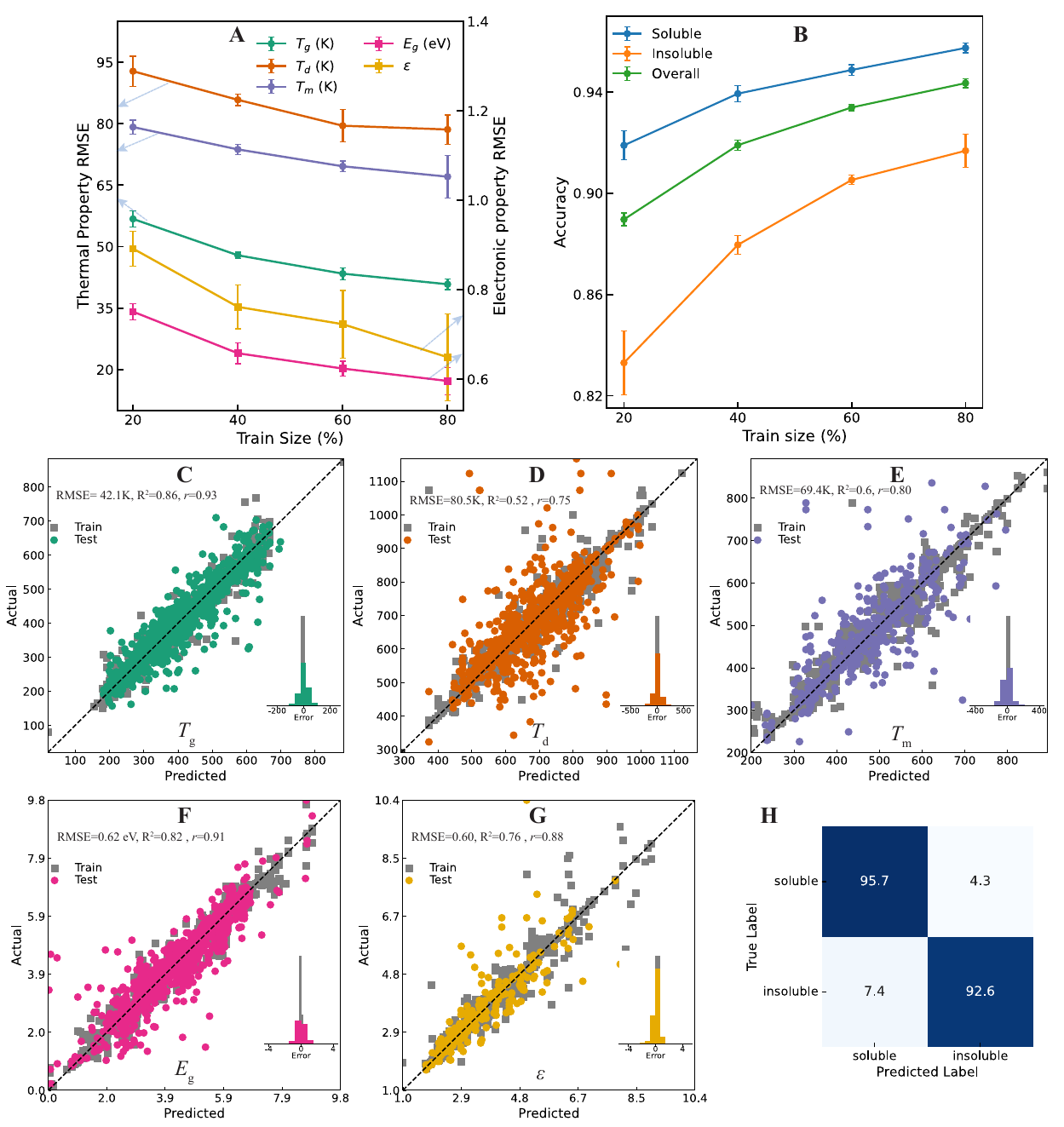} 
        \caption{\textbf{Performance of fine-tuned \textsc{polyT5}-medium models for various property prediction tasks.}
        (\textbf{A}) Learning curves showing the root-mean-square errors (RMSEs) for thermal and electronic property predictions.
        (\textbf{B}) Learning curves for solubility prediction. For panels \textbf{A} and \textbf{B}, error bars represent the standard deviation from five different random train-test splits.
        (\textbf{C--G}) Parity plots for a representative split, comparing predicted and experimental values for glass transition temperature ($T_{\rm g}$), thermal decomposition temperature ($T_d$), melting temperature ($T_m$), electronic band gap ($E_g$), and dielectric constant ($\varepsilon$).
        (\textbf{H}) Confusion matrix for a representative split for solubility prediction, with values reported in percentage.
        }
	\label{fig4}
\end{figure}

All three \textsc{polyT5} base model variants were fine-tuned to predict thermal ($T_{\rm g}$, $T_{\rm d}$, $T_{\rm m}$), electronic ($E_{\rm g}$, $\varepsilon$), and solubility properties.\cite{methods} The mean absolute error (MAE) values across epochs for the \textsc{polyT5}-medium model, evaluated on thermal and electronic properties, are presented in Figure~\ref{figS4}A–C. As expected, the MAE decreased with increasing epochs, saturating around 30 epochs, and also showed a general decrease with larger training set sizes. For the \textsc{polyT5}-medium model, learning curves showing the RMSE values on unseen test sets are presented in Figure~\ref{fig4}A, while the corresponding plots for $R^2$ and Pearson correlation coefficient ($r$) are shown in Figure~\ref{figS4}D-E. Error bars represent the standard deviation obtained from five different random train-test splits. As expected, the RMSE values decrease, and both $R^2$ and $r$ increase with larger training set sizes. The average RMSEs for an 80\% training size on the unseen test set were 40.8, 78.6, and 67.1 for $T_{\rm g}$, $T_{\rm d}$, and $T_{\rm m}$, respectively. For $E_{\rm g}$ and $\varepsilon$, the RMSE values were 0.596 and 0.649, respectively. The parity plots for each property for a representative split with an inset of the error distribution are shown in Figure~\ref{fig4} (C-G). As shown in the parity plots, the predicted values closely align with the true values for all properties. The corresponding error distributions are centered around zero with relatively small spreads, indicating good model performance.

For the solubility of polymers across various solvents, the learning curves for the prediction accuracy of the soluble, insoluble and overall cases are presented in Figure~\ref{fig4}B, with error bars representing the standard deviation between five random splits. As expected, accuracy improves with increasing training set size. For an 80\% training size, the accuracies for soluble and insoluble cases reached 0.957 and 0.917, respectively, resulting in an overall accuracy of 0.943. The confusion matrix for a representative split, shown in Figure~\ref{fig4}H, further illustrates this performance, with 95.7\% of soluble and 92.6\% of insoluble cases correctly classified.

The results for \textsc{polyT5}-small and \textsc{polyT5}-large are shown in Figures \ref{figS5} and \ref{figS6}, respectively. Overall, the fine-tuned \textsc{polyT5}-small models underperformed compared to the medium model, while the \textsc{polyT5}-large models showed marginal improvements over the medium variant. Considering the balance between predictive accuracy and computational cost, the \textsc{polyT5}-medium model was selected for subsequent analyses.

\subsection*{Downstream Task 2: Generative design}
For the design of dielectric materials, the $T_{\rm g}$ of polymers plays a critical role in determining their thermal and mechanical stability under operating conditions.\cite{lia23:2777} Given that the corresponding dataset is the largest in our collection and encompasses a broad range of chemical diversity, it was selected for the generative design task. The \textsc{polyT5} models were fine-tuned\cite{methods} on the $T_{\rm g}$ dataset, where $T_{\rm g}$ values were provided as input and the corresponding PSELFIES strings served as output (Figure~\ref{fig2}E). The fine-tuned models were subsequently employed to generate candidate polymers targeting specific $T_{\rm g}$ values.

During generation, several hyperparameters influence the selection of the next token and, consequently, the resulting PSELFIES. These include the number of fine-tuning epochs, the T5 temperature parameter (which scales the logits to adjust randomness or confidence in predictions), and nucleus sampling (\texttt{top\_p}), which limits token selection to the smallest possible set whose cumulative probability meets a specified threshold. To systematically assess how these hyperparameters govern the generation of hypothetical polymers for small, medium and large-\textsc{polyT5} models, we varied the number of training epochs from 1 to 15, the temperature from 0.1 to 2.0 in increments of 0.1, and used \texttt{top\_p} values of 0.75 and 0.95, generating 10,000 polymers for each configuration.

To assess the performance and quality of the generated hypothetical polymers, four evaluation metrics were employed. First, SMILES Validity (SV) was determined using RDKit\cite{rdkit} to ensure the chemical validity of the generated structures. Next, Training Set Deduplication (TSD) filtered out any candidates that were already present in the training dataset. The Dataset Deduplication (DD) step removed duplicates within the generated set itself, retaining only unique candidates. Finally, PSMILES Validity (PV) ensured that each retained candidate contained exactly two Astatine (At) atoms, each with a valency of one, as required by the polymer design rules. These filters follow a nested relationship: SV~$\supset$~TSD~$\supset$~DD~$\supset$~PV, such that polymers passing the PV criterion are valid, unique, and novel structures generated by the \textsc{polyT5} models.

\begin{figure} 
	\centering
	\includegraphics[width=1.0\textwidth]{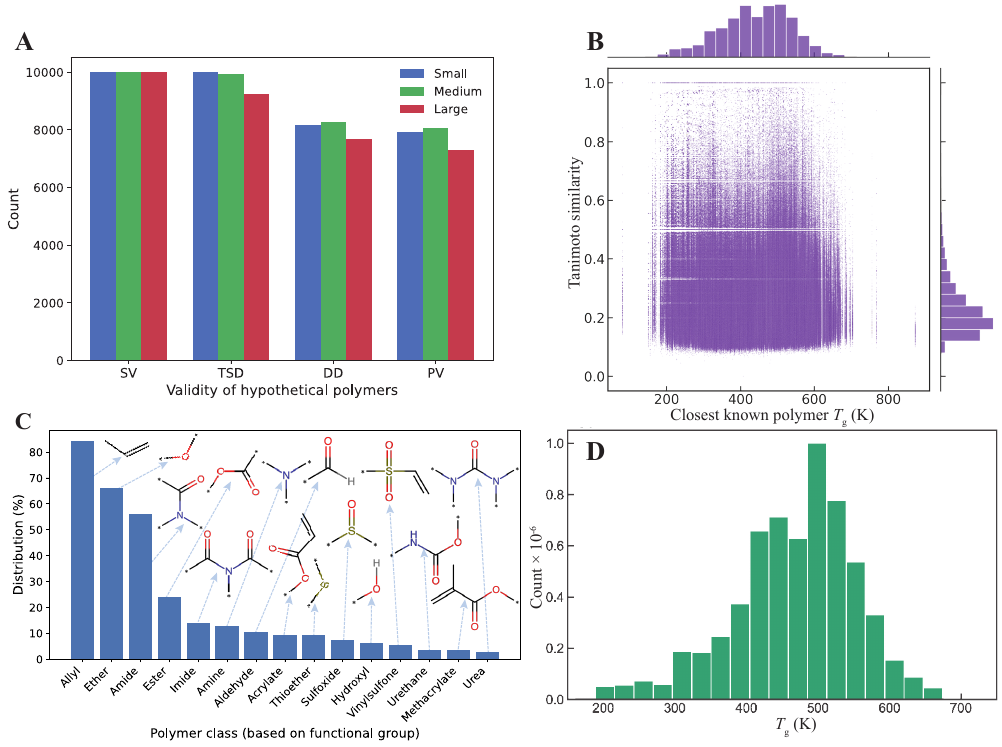} 
\caption{\textbf{Hypothetical candidate generation using \textsc{polyT5}.}
        (\textbf{A}) Optimal combination of fine-tuning epoch, T5 sampling temperature, and sampling probability identified for small (14, 0.9, 0.75), medium (6, 1.1, 0.75), and large (8, 1.1, 0.95) \textsc{polyT5} models for maximizing the generation of valid hypothetical polymers. SV, TSD, DD, and PV represent successive validation filters: validity of SMILES strings verified by RDKit (SV), removal of duplicates against the training set (TSD), deduplication within the generated hypothetical polymer dataset (DD), and polymer validity (PV) ensuring the presence of exactly two Astatine (At) atoms each with valency one. Note that PV~$\subset$~DD~$\subset$~TSD~$\subset$~SV.
        (\textbf{B}) Distribution of Tanimoto similarity values between each of the over 6 million \textsc{polyT5}-generated hypothetical polymers and their closest known polymer in the training dataset, as a function of the closest polymer’s $T_{\rm g}$ (K).
        (\textbf{C}) Distribution of polymer classes based on the presence of functional groups in the 6 million generated hypothetical polymers.
        (\textbf{D}) Distribution of predicted $T_\mathrm{g}$ values by \textsc{polyT5}-medium for over 6 million candidate polymers.
        }
	\label{fig3}
\end{figure}

Figure~\ref{fig3}A presents the performance of hypothetical candidate generation by the \textsc{polyT5}-small, medium, and large models targeting a $T_{\rm g}$ of 500 K, using the optimal combination of epochs, sampling temperature, and \texttt{top\_p} values that maximize the number of candidates passing the PV filter. The performance across all possible combinations of these hyperparameters is provided in Figure~\ref{figS1}. Our results revealed important trends in how fine-tuning epoch and generation hyperparameters influence the quality of hypothetical polymer candidates. As the number of fine-tuning epochs increased, the models initially improved, producing fewer invalid candidates. However, beyond a certain point, extended fine-tuning began to increase the number of duplicates within the generated set — a trend especially pronounced for the medium and large models. Higher T5 temperatures increased the likelihood of generating invalid PSMILES, while lower temperatures led to excessive duplication, either producing identical hypothetical polymers or reproducing structures from the training set. The effect of \texttt{top\_p} was also notable: a higher value (0.95) allowed a broader range of tokens during sampling, resulting in greater chemical diversity but also a higher fraction of invalid structures. In contrast, a lower \texttt{top\_p} (0.75) restricted token choices, reducing invalid outputs but increasing duplication. These observations highlight the delicate trade-offs involved in tuning candidate generation and emphasize the need for a careful balance between model training and generation parameters.


In the process of varying fine-tuning epochs and generation hyperparameters, a total of 6,171,066 valid candidate polymers were generated that passed the PV filter. To assess the novelty and structural similarity of these hypothetical polymers relative to the training set, Tanimoto similarity was computed between each generated polymer and the known polymers in the training data, using ECFP6 2048-bit fingerprints as implemented in RDKit.\cite{rdkit} To eliminate terminal effects and better approximate infinite polymer chains, the two ends of each repeating unit were connected to form a loop prior to fingerprint calculation. Figure~\ref{fig3}B presents the $T_{\rm g}$ of the most similar (highest Tanimoto similarity) known polymer in the training set against the corresponding similarity value. As expected, most generated polymers are close in $T_{\rm g}$ to 500~K, consistent with the target value specified during generative design. Importantly, the distribution of Tanimoto similarity values indicates that while generated polymers are guided by examples in the training set, their molecular structures are largely distinct. Further insight into the chemical diversity of the generated polymers is provided in Figure~\ref{fig3}C, which shows the distribution of functional groups present in the generated hypothetical candidates, with a detailed breakdown listed in Table~\ref{tabS4}. As shown, allyl, ether, and amide functional groups dominate the generated set, while a sizable portion also contains ester, imide, amine, aldehyde, acrylate, and thioether groups. This highlights the model’s ability to capture underlying structure–property relationships while generating a chemically diverse set of novel polymer candidates tailored to the desired thermal property.

The fine-tuned \textsc{polyT5}-medium model was employed to predict the $T_{\rm g}$ values of all 6 million generated hypothetical polymers, with the resulting distribution shown in Figure~\ref{fig3}D. As expected, the distribution is centered around the target value of 500~K, although it exhibits a slight skew toward lower $T_{\rm g}$ values, likely due to the higher abundance of training samples below 500~K relative to those above (see Figure~\ref{prop_train}). To further assess the model’s generative capability, an additional set of candidates was generated using \textsc{polyT5}-medium with a target $T_{\rm g}$ of 300~K. The predicted $T_{\rm g}$ distribution for these candidates, shown in Figure~\ref{fig300K}, now peaks near 300~K as anticipated. These results collectively demonstrate that \textsc{polyT5} has effectively learned the intrinsic relationship between polymer chemical structure and $T_{\rm g}$, enabling the generation of candidates tailored to specified target temperatures.

\subsection*{Impact of pre-training: Ablation study on \textsc{polyT5}}
An ablation study was conducted using the \textsc{polyT5}-medium model to evaluate the impact of pre-training on both property prediction and candidate generation tasks. Table \ref{tabS5} summarizes prediction results for thermal, electronic, and solubility properties. Across all cases, fine-tuning the pre-trained model significantly outperformed training from randomly initialized weights, despite using the same architecture and training protocol. For example, in the case of $T_{\rm g}$, the model trained without pre-training achieved an average RMSE of 89.35,K and $R^2$ of 0.31 across five data splits, while the pre-trained model reduced the RMSE to 40.82,K and improved $R^2$ to 0.86. Figures \ref{abl_gen_part1} and \ref{abl_gen_part2} present the generation results for hypothetical polymers targeting a $T_{\rm g}$) of 500~K. The pre-trained model consistently produced a higher number of valid candidates with the desired property (500±50~K), and demonstrated reduced sensitivity to fine-tuning epochs and T5 temperature. Notably, the SELFIES reproducibility (SR) metric—defined as the fraction of generated SELFIES strings that, when converted to SMILES and back to SELFIES, yield the identical string—showed a 5-fold improvement with pre-training. Since SR measures the model’s ability to produce canonical SELFIES representations, it reflects a particularly challenging aspect of chemical language modeling. These findings highlight the critical role of large-scale pre-training on 100 million polymers, which equips \textsc{polyT5} with an in-depth understanding of structural features in chemical representations and facilitates effective transfer learning for property prediction and polymer generation.

\subsection*{Dielectric polymer design}
In designing dielectric polymers, we considered not only a high $\epsilon$ ($>$3) but also key properties such as a wide $E_{\rm g}$ ($>$4 eV) and high $T_{\rm g} (>$400~K) to ensure thermal and electrical stability. Practical processing criteria, including melt processability ($T_{\rm m}$-$T_{\rm g}$$>$100~K, $T_{\rm d}$-$T_{\rm g}$$>$100~K) and solubility either in water or ethanol, were also enforced to ensure that the selected candidates could be synthesized and processed experimentally. These thresholds collectively balance performance, stability, and manufacturability, ensuring that screened polymers are suitable for high-performance dielectric applications.

\begin{figure} 
	\centering
	\includegraphics[width=0.95\textwidth]{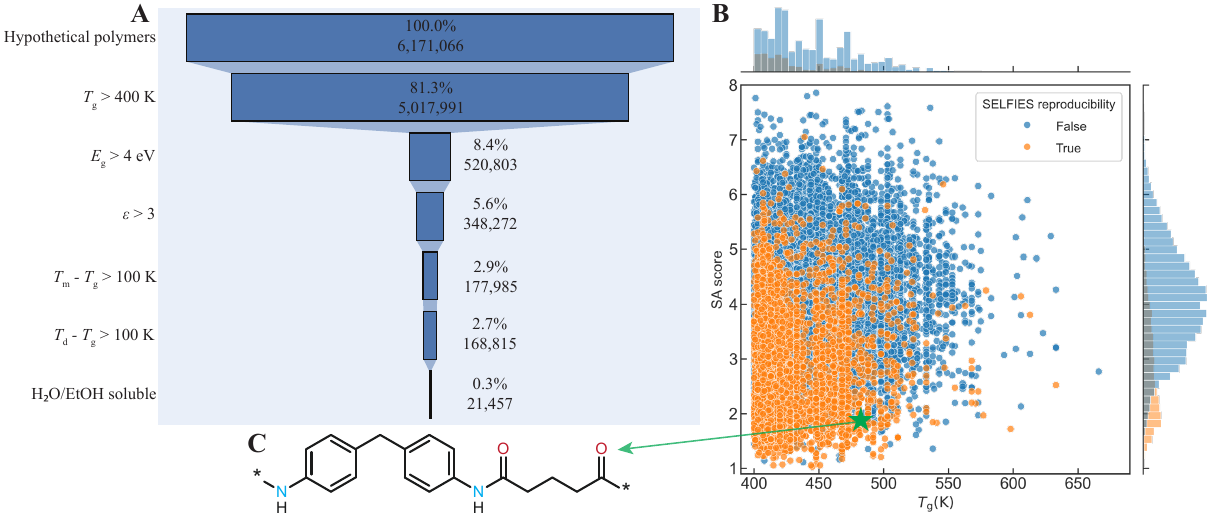} 
        \caption{
        \textbf{Designing dielectric polymers with \textsc{polyT5}.}
        (\textbf{A}) Screening of candidates based on thermal, electronic, and solubility criteria.
        (\textbf{B}) Synthetic accessibility scores of selected polymers versus predicted $T_\mathrm{g}$, annotated by SELFIES reproducibility, with corresponding marginal histograms.
        (\textbf{C}) Selected polymer for experimental validation
        }
	\label{fig5}
\end{figure}

To identify promising candidates for high-performance dielectric applications, a multi-step screening strategy was applied (see Figure~\ref{fig5}A) to the 6,171,066 hypothetical polymers generated by the \textsc{polyT5} models. First, general property criteria were used to narrow down the set: polymers with a predicted $T_{\rm g}$ exceeding 400~K accounted for approximately 5 million candidates. Of these, 520,803 exhibited a predicted $E_{\rm g}$ greater than 4~eV, and 348,272 satisfied the target dielectric constant range. Subsequently, 177,985 candidates met the melt processing requirement ($T_{\rm m} - T_{\rm g} > 100$~K), while 168,815 satisfied the thermal stability condition ($T_{\rm d} - T_{\rm g} > 100$~K). Finally, solubility criteria were applied to prioritize polymers compatible with common, low-cost, and environmentally friendly solvents. Specifically, water and ethanol were selected due to their status as widely available, green solvents\cite{ald16:3879} frequently used in industrial and laboratory processing. This final filter identified 21,457 polymers predicted to be soluble in at least one of these solvents. This systematic, multi-property screening workflow enabled the identification of a focused set of candidates that combine desirable dielectric, thermal, and processing attributes, suitable for further experimental validation.

To evaluate the synthetic accessibility (SA) of the generated hypothetical polymers, the SA score for each candidate was calculated by first replacing the placeholder atoms ([At] or [*]) with hydrogen atoms, effectively treating each polymer as a monomer. The SA scores were then computed using RDKit\cite{rdkit}. As shown in Figure~\ref{figS7}, nearly all known polymers in the training set exhibited SA scores below 6, with most falling within the 2–3 range. Figure~\ref{fig5}B presents the distribution of predicted $T_{\rm g}$ values against SA scores for the screened candidates, with side histograms and annotations based on the SR metric, effectively testing whether \textsc{polyT5} generated SELFIES strings in a chemically consistent and robust manner. Among the 21,457 screened candidates, 3,978 were found to pass the SR test. Notably, candidates with reproducible SELFIES strings exhibited lower SA scores on average, while those failing the reproducibility test tended to have higher SA scores, including 535 candidates with values exceeding 6. These findings suggest that polymers with reproducible SELFIES representations are generally less structurally complex, making them easier for the model to learn and, in turn, more synthetically accessible.

\subsection*{Experimental validation}
The polymer shown in Figure~\ref{fig5}C was selected for experimental verification owing to its ease of synthesis. The polymer was prepared by solution polycondensation using glutaryl dichloride and 4,4'-diaminodiphenylmethane under basic conditions. $^{1}$H NMR spectroscopy (Figure~\ref{exp_results_SI}) confirmed the expected chemical structure and composition. The measured and predicted properties are summarized in Table~\ref{tabS6}. The experimental $T_{\rm g}$ was 472~K, in good agreement with the predicted value of 483~K. The $T_{\rm m}$ appeared at 543~K, about 60~K lower than predicted, suggesting semi-crystalline domains. A higher $T_{\rm d}$ was observed at 607~K, 36~K less than the predicted value. From DFT calculations, the $E_{\rm g}$ was determined to be 4.53~eV, showing remarkable consistency with the predicted 4.45~eV. All deviations fall within the models’ error ranges. Details of the synthetic procedures, additional characterization, and DFT calculations are provided in the Supporting Information.\cite{methods}. 

\subsection*{Agentic AI framework: Integration of LLM and \textsc{polyT5} models}
To demonstrate how such a polymer generative capability can have enhanced accessibility, we developed an agentic AI framework that integrates a general-purpose large language model (\textit{gpt-5-nano}\cite{OpenAI_gpt5_nano}) with fine-tuned \textsc{polyT5} models under a single conversational interface.\cite{methods} This framework enables users to query thermal, electronic, solubility, and dielectric properties of polymers, as well as perform generative design tasks, using natural language. All inputs and outputs are expressed in SMILES format, with internal conversion to and from SELFIES for inference. During generation, invalid candidates are automatically filtered, ensuring that only valid polymer SMILES are returned. A schematic overview (Figure~\ref{chatbot}) illustrates how user queries are parsed by the controller LLM, routed to the relevant \textsc{polyT5} model, and returned as validated outputs.  

The framework unifies multiple models under one platform, removing the need to switch between tools, while natural language interaction broadens accessibility to non-experts. It further ensures robustness through automated input handling, reproducibility via standardized schemas, and extensibility through a modular design that allows seamless integration of new models. Together, this agentic AI framework lowers the barrier to advanced polymer modeling, combining the reasoning capabilities of LLMs with the predictive and generative power of \textsc{polyT5}.

\section*{Discussion}
In this work, we introduced \textsc{polyT5}, the first foundation large language model (LLM) tailored for polymers. Trained on over 100 million polymer structures in SELFIES representation with domain-specific tokens using the T5 architecture, \textsc{polyT5} was developed in three variants of varying depth and embedding dimensions and fine-tuned for two key downstream tasks: (a) property prediction across thermal, electronic, and solubility properties, and (b) generative design of polymers with targeted glass-transition temperatures ($T_{\rm g}$).

For property prediction, \textsc{polyT5} achieved RMSEs of 40.82, 67.07 and 78.59~K for $T_{\rm g}$, $T_{\rm m}$, and $T_{\rm m}$, respectively, and 0.60~eV and 0.65 for $E_{\rm g}$ and $\epsilon$. For polymer solubility classification, the model reached an overall accuracy of 0.94, with soluble and insoluble cases predicted with accuracies of 0.96 and 0.92, respectively. In the generative design task, \textsc{polyT5} successfully produced hypothetical polymers with targeted $T_{\rm g}$ values of 300 and 500~K, with predicted distributions centered near the specified targets. Tanimoto similarity analysis confirmed that the generated candidates were chemically diverse rather than trivial replicas, demonstrating that the model captures structure–property relationships beyond simple memorization.

To demonstrate the practical utility of the developed framework, we applied \textsc{polyT5} to the design of high-energy dielectric polymers. Leveraging its generative and predictive capabilities, over 6 million candidates were generated and subsequently screened with property-prediction models. Using cutoff criteria of $\epsilon\geq$3, $E_{\rm g}\geq$4~eV, $T_{\rm g}\geq$400~K, along with melt processability and solubility requirements, more than 20,000 candidates were identified as promising. One representative polymer was synthesized and experimentally validated, showing good agreement between measured and predicted properties, while additional DFT calculations confirmed the predicted band gap; together, these validations further support the reliability of the framework.

Beyond model development and validation, we integrated \textsc{polyT5} within an agentic AI framework that combines it with a general-purpose LLM, enabling natural language interaction for property prediction and generative design. This conversational interface lowers technical barriers by handling input validation, format conversion, and model selection automatically, thereby making advanced polymer modeling accessible to both experts and non-experts. Taken together, these contributions highlight the promise of domain-specific foundation models in polymer science, \textsc{polyT5} captures polymer structure–property relationships and extends this knowledge to prediction and generative design. With fewer than 7.5 million parameters, it achieves both accuracy and efficiency, establishing a practical, accessible, and extensible foundation for accelerated polymer discovery and future applications to more complex materials systems.



\newpage
\begin{table} 
	\centering
        \caption{\textbf{Properties of the selected polymer as obtained from \textsc{polyT5}, experiment, and DFT.} Predictions were made using the \textsc{polyT5}-medium model. The band gap ($E_{\rm g}$) was computed using DFT, while the remaining properties were measured experimentally.}
	\label{tabS6} 
        \begin{tabular}{lccccc}
        \hline
        \textbf{Property} & $T_{\rm g}$ (K) & $T_{\rm m}$ (K) & $T_{\rm d}$ (K) & $E_{\rm g}$ (eV) \\ 
        \hline
        \textsc{polyT5} & 483 & 603 & 643 & 4.45 \\ 
        Experiment/DFT & 472 & 543 & 607 & 4.53\\
        \hline
        \hline
        \end{tabular}
\end{table}

%


\clearpage 

%
\bibliography{polyT5} 
\bibliographystyle{sciencemag}

%
%
%
%
%
%


\section*{Acknowledgments}
\paragraph*{Funding:}
The authors acknowledge financial support by the Office of Naval Research through grants
N00014-19-1-2103 and N00014-20-1-2175.

\paragraph*{Author contributions:}
H.~S. was the primary architect of the project, including dataset development, base and fine-tuned models, candidate generation and screening workflow, and manuscript preparation. A.~S. assisted in developing the base model. S.~S. generated the hypothetical candidate dataset using known reactions employed for pre-training the base model. W.~X. conducted the experimental validations and contributed to writing the corresponding sections. R.~R. conceived the project, provided overall guidance, and supervised the work.

\paragraph*{Competing interests:}
``There are no competing interests to declare.''
\subsection*{Supplementary materials}
Materials and Methods\\
Figs. S1 to S13\\
Tables S1 to S5\\


\newpage


\renewcommand{\thefigure}{S\arabic{figure}}
\renewcommand{\thetable}{S\arabic{table}}
\renewcommand{\theequation}{S\arabic{equation}}
\renewcommand{\thepage}{S\arabic{page}}
\setcounter{figure}{0}
\setcounter{table}{0}
\setcounter{equation}{0}
\setcounter{page}{1} 


\begin{center}
\section*{Supplementary Materials for\\ \scititle}


Harikrishna Sahu$^{1}$,
Wei Xiong$^{1}$,
Anagha Savit$^{1}$,
Shivank S Shukla$^{1}$,
Rampi Ramprasad$^{1\ast}$\\ 

\small$^{1}$School of Materials Science and Engineering, Georgia Institute of Technology, Atlanta, GA 30332, USA\\
\small$^\ast$Corresponding author. Email: rampi.ramprasad@mse.gatech.edu\\
\end{center}

\subsubsection*{This PDF file includes:}
Materials and Methods\\
Figures S1 to S13\\
Tables S1 to S5\\


\newpage


\subsection*{Materials and Methods}



\subsubsection*{Generation of reliable hypothetical polymers for pre-training}
Over 100 million of hypothetical homopolymers were generated using well-established polymerization reactions, encompassing common polymer chemistries such as polyamides, polyimides, polyesters, polyethers, polyureas, and polyurethanes. These candidates were created by reacting commercially available small molecules sourced from databases including eMolecules, ChEMBL, and ZINC-15 via known polycondensation reactions, following strategies similar to those reported in previous studies.\cite{odi04:39, ste15:2324, gau16:D945, eMolecules, gur24:6107, kim23:318, ohn23:5539, yue24:2465} To further enhance structural diversity, the ring-opening metathesis polymerization (ROMP) was employed as demonstrated in prior work.\cite{odi04:544, gur24:6107} Several click reactions were also considered, such as Cu-catalyzed azide–alkyne cycloaddition (CuAAC), strain-promoted azide–alkyne cycloaddition (SPAAC), thiol–ene/yne/bromo coupling, Diels–Alder (furan–maleimide), sulfur fluoride exchange (SuFEx), and oxime-based click reactions.\cite{kl01:2004, gen21:963, hoy10:1540, low10:4745, zha15:97, gan13:1, wan17:11203, col16:2581} These approaches significantly expanded the chemical and functional space of the generated polymer dataset.

\subsubsection*{Tokenizer vocabulary}
For tokenizing the PSELFIES strings, each substring enclosed within square brackets (e.g., [C], [O]) was treated as a distinct token,\cite{ana25:04233, ind24:12348} resulting in a base vocabulary of 199 unique tokens. Several special tokens were also introduced, including start- and end-of-sequence markers, unknown and padding tokens, a whitespace marker, and 100 sentinel tokens for masking during pre-training. To further expand the vocabulary and enable property-conditional generation and prediction, an additional 154 tokens were incorporated. These included property names, numerical digits (0–9), decimal point (.), units, arithmetic and relational operators (+, -, $>$, $<$, =, etc.), boolean values, and a set of common polymer-related keywords. This resulted in a final vocabulary size of 458 tokens. To ensure compatibility with the SentencePiece\cite{kud18:06226} tokenizer framework, all SELFIES tokens and additional custom tokens were included as predefined tokens.

\subsubsection*{Model pre-training: Architectures, span Masking, and optimization}
We pre-trained three variants of a polymer-specialized T5-based\cite{col23:10683} model: \textsc{polyT5}-small, \textsc{polyT5}-medium, and \textsc{polyT5}-large, using a masked span prediction objective within a sequence-to-sequence framework. Polymer structures were represented using SELFIES strings and tokenized using a custom SentencePiece\cite{kud18:06226} tokenizer comprising 199 unique tokens corresponding to the SELFIES vocabulary. Each model variant follows the standard T5 architecture but differs in size and complexity (see Table \ref{tabS0}): \textsc{PolyT5}-small uses an embedding dimension (\texttt{d\_model}) of 128 with 3 encoder and decoder layers ($\sim$1.44 million parameters); \textsc{PolyT5}-medium uses \texttt{d\_model} = 256 with 4 layers ($\sim$7.46 million parameters); and \textsc{PolyT5}-large uses \texttt{d\_model} = 512 with 8 layers ($\sim$58.98 million parameters). All models employ relative positional encodings with a maximum input length of 200 tokens.

The training objective follows the span corruption strategy introduced in the original T5 model. For each polymer sequence, up to 8 non-overlapping masked spans (each up to 3 tokens long) were randomly selected to mask up to 15\% of the input tokens. These spans were replaced with sentinel tokens (\texttt{$<$extra\_id\_n$>$}) in the input sequence, and the target sequence was constructed by concatenating the masked spans, each prefixed with its corresponding sentinel token. The sentinel tokens were assigned in increasing numerical order of $n$ and placed such that no two masked spans were adjacent, ensuring at least one unmasked token between them.

The models were trained on $\sim$90 million masked polymer sequences (90\% of the dataset), using a batch size of 450 for up to 5 epochs on a single NVIDIA L40S GPU. The remaining 10\% was reserved for validation and testing. Training loss was calculated using the token-level cross-entropy objective and optimized with the AdamW optimizer. Batch-level training loss was monitored and logged, with checkpoints saved after each epoch. After training, the model weights were saved for subsequent fine-tuning and inference, while a single, pre-defined tokenizer was used throughout.

\subsubsection*{Downstream Task 1: Property Prediction}
To fine-tune \textsc{polyT5} for polymer property prediction, we formulated the task as a sequence-to-sequence problem. The model input consisted of polymer structures encoded as SELFIES strings, with supplementary information provided when relevant. For instance, dielectric constant prediction included frequency information in log scale, and solubility prediction incorporated the SELFIES representation of solvents prefixed appropriately. The target output was either a continuous value for thermal or electronic properties, or a categorical label (e.g., “soluble” or “insoluble”) for classification tasks such as solubility.

Fine-tuning was performed separately for each property using data splits ranging from 20\% to 80\% for training and the remainder for testing. Five different random splits were employed to generate learning curves and assess model robustness. Both input and target sequences were tokenized using a single, pre-defined SentencePiece tokenizer, with sequences truncated or padded to a maximum length of 200 tokens.

Training proceeded for up to 30 epochs with a batch size of 16, using the AdamW optimizer with a learning rate of 3e$^{-4}$ and a weight decay of 0.01 for regularization. The loss function was token-level cross-entropy, where padding tokens in the target sequences were replaced by –100 to be ignored during loss calculation. Evaluation was conducted at the end of each epoch, measuring mean absolute error (MAE) between predicted and true property values. For regression tasks, predictions were generated using beam search with a beam width of 4, decoded into floating-point numbers, and filtered to remove any invalid or non-numeric outputs.\\

\textbf{Model input-output examples for property prediction tasks}:
\begin{enumerate}
\item \textbf{Thermal and electronic Properties} ($T_{\rm g}$, $T_{\rm m}$, $T_{\rm d}$, $E_{\rm g}$):\\
\textit{INPUT:} PSELFIES\\
\textit{Example:} \texttt{[C][C][C][C][Branch1][C][At][C][At]}\\
\textit{OUTPUT:} Property value, e.g., \texttt{236.0}

\item \textbf{Dielectric constant ($\varepsilon$):}\\
\textit{INPUT:} Property tag with log(frequency) followed by PSELFIES\\  
\textit{Example:} \texttt{property 4.1; polymer [C][C][Branch1][C][At][C][At]}\\  
\textit{OUTPUT:} Property value, e.g., \texttt{3.7}

\item \textbf{Polymer solubility:}\\  
\textit{INPUT:} Polymer and solvent represented in SELFIES format\\
\textit{Example:} \texttt{polymer [C][C][Branch1][C][At][C][At]; solvent [C][C][C][O][C]}
\texttt{[Ring1][Branch1]}\\  
\textit{OUTPUT:} Classification token, e.g., \texttt{soluble} or \texttt{insoluble}
\end{enumerate}

\subsubsection*{Downstream Task 2: Polymer Generation}
For the polymer generation task, we fine-tuned the pre-trained \textsc{polyT5} models to generate complete polymer structures in SELFIES format, conditioned on a target property. In this study, $T_{\rm g}$ was selected as the target property due to its significance in materials design, broad coverage across diverse chemistries, and the relative completeness of the available data in the training set. This task was formulated as a conditional sequence generation problem, where the model learns to map scalar property values to valid polymer sequences.

The fine-tuning dataset consisted of paired ($T_{\rm g}$, PSELFIES) examples, with 90\% of the data used for training and 10\% for validation. All sequences were tokenized using the same SentencePiece tokenizer employed during pre-training and padded to a maximum sequence length of 200 tokens. No masking was applied during this phase, as the model was trained to generate the full target sequence autoregressively, conditioned on the input property string.

Fine-tuning was performed using the HuggingFace Seq2SeqTrainer API for up to 15 epochs, with a batch size of 16. The training used the AdamW optimizer with a learning rate of 3e$^{-4}$ and a weight decay of 0.01. Training, evaluation, and checkpointing were performed at the end of each epoch. The objective was to minimize the token-level cross-entropy loss, ignoring padding tokens in the target sequence using a special –100 label. Generation was enabled during evaluation, with a maximum output length of 200 tokens. Upon completion, the fine-tuned models and their associated tokenizers were saved for downstream inference and sampling tasks.

During inference, polymer sequences were generated using a sampling-based approach instead of beam search (which was used for the property prediction task). To systematically assess the effects of temperature and sampling diversity on generation quality, we evaluated models fine-tuned for 1 to 15 epochs using combinations of \texttt{top\_p} $\in$ \{0.75, 0.95\} sampling thresholds and temperature values ranging from 0.1 to 2.0 in increments of 0.1. This setup enabled a comprehensive exploration of generation behavior under varying stochastic sampling conditions.\\

\textbf{Hypothetical polymer generation based on target $T_{\rm g}$}:
\begin{itemize}
   \item
    \textit{INPUT:} Target property value (e.g., \texttt{236.0})\\
    \textit{OUTPUT:} Corresponding polymer in SELFIES format\\
    \textit{Example:} \texttt{[C][C][C][C][Branch1][C][At][C][At]}
\end{itemize}

\subsubsection*{Agentic AI Framework}
An agentic AI framework was constructed by integrating a general-purpose LLM (\textit{gpt-5-nano}\cite{OpenAI_gpt5_nano}) with task-specific polymer models. The LLM acted as the controller agent, implemented via the PydanticAI library, which provides structured tool invocation and schema validation. Each \textsc{polyT5} model (thermal predictors, Eg predictor, solubility predictor, and Tg-conditioned generator) was wrapped as a PydanticAI Tool with standardized input and output schemas defined using Pydantic BaseModels.

The supervisor LLM was supplied with routing rules in its system prompt to ensure that user queries were directed to the appropriate tool. All inputs and outputs were expressed as SMILES strings: prediction tasks required polymer SMILES with two [*] termini, while solubility additionally required solvent SMILES without termini, and generation required only a target Tg value. In the backend, SMILES were automatically converted to SELFIES for model inference. For generation, candidate SELFIES were converted back to polymer SMILES, with invalid strings rejected to ensure that only valid polymer SMILES were returned. All scientific predictions and generations were executed exclusively by the \textsc{polyT5} models.

A lightweight Streamlit interface exposed the framework as a conversational chatbot. User queries were relayed to the supervisor LLM, which invoked the relevant tool and returned outputs displayed as text. The interface retained chat history within each session, allowing the dialogue to proceed across successive queries.

\subsubsection*{First-Principles Calculations}
The initial infinite-chain polymer structure was generated using the PSP\cite{sah22:2737} package, with vacuum regions of 12\AA\ along the non-periodic directions to minimize interchain interactions. Density functional theory (DFT) calculations were then performed using the Vienna Ab initio Simulation Package (VASP)\cite{kre96:15}, employing the Perdew–Burke–Ernzerhof (PBE) exchange–correlation functional\cite{per96:3865} and a plane-wave energy cutoff of 400 eV. Geometry optimization was considered converged when the maximum force on any atom was below 0.01 eV/\AA. The optimized geometries were subsequently used to compute the electronic structure using the HSE06 hybrid functional.\cite{hey03:8207}

\subsubsection*{Experimental validation}
\textbf{Synthesis of PA from Glutaryl dichloride and 4,4'-Diaminodiphenylmethane}: All glassware was oven-dried and cooled under a nitrogen atmosphere before use. A three-neck round-bottom flask was fitted with a dropping funnel, condenser, and magnetic stir bar, and nitrogen flow was maintained throughout the reaction. Glutaryl dichloride (1.194~g, 7.07~mmol, 1.0~equiv) and 4,4'-diaminodiphenylmethane (1.400~g, 7.07~mmol, 1.0~equiv) were dissolved together in 200~mL of anhydrous dichloromethane (DCM) and cooled in an ice–water bath to 0–5\textdegree C. Triethylamine (0.785~g, 7.78~mmol, 1.1~equiv) was added dropwise over about 30 min under vigorous stirring, keeping the temperature below 10\textdegree C to control the heat release. After TEA addition, the mixture was allowed to reach room temperature ($\sim$25\textdegree C) and stirred for an additional 2~h. A white precipitate appeared, indicating polymer formation. The crude product was collected by filtration, washed several times with cold methanol/diethyl ether (1:1 v/v) to remove TEA hydrochloride and unreacted monomers, and then dried in a vacuum oven at 40\textdegree C for 12~h to give the purified PA.\\

\textbf{Characterization}: The chemical structure and composition of the polymers were confirmed by $^{1}$H nuclear magnetic resonance ($^{1}$H NMR) spectroscopy in {d6-DMSO} at room temperature. Characteristic peaks corresponding to both monomer units were clearly observed in the $^{1}$H NMR spectra (See Figure~\ref{exp_results_SI}).The $T_{\rm g}$ was measured using differential scanning calorimetry (DSC) under a nitrogen atmosphere with a heating rate of 10\textdegree C/min.


\clearpage




\newpage




\begin{figure}
	\centering
	\includegraphics[width=0.5\textwidth]{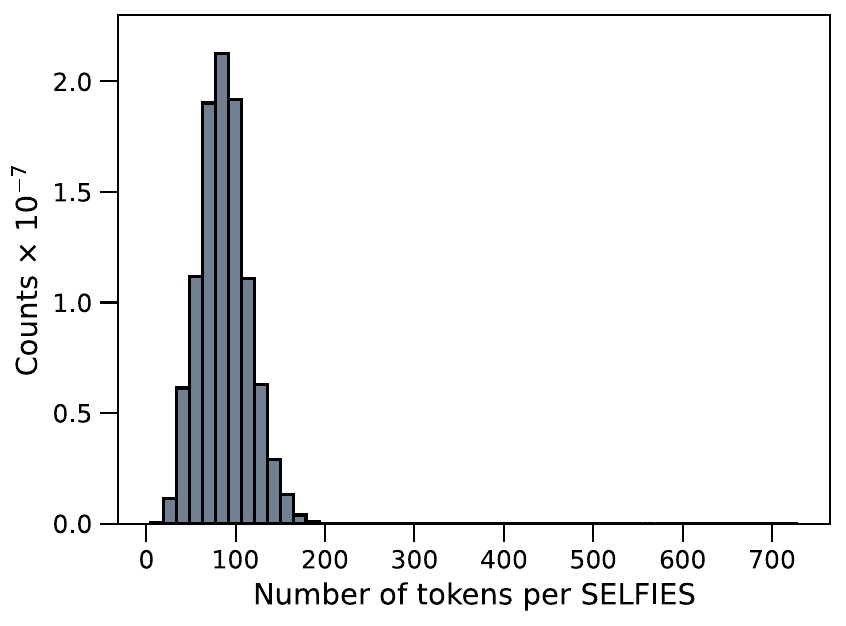}
        \caption{
        \textbf{Distribution of token counts per SELFIES string.}
        The histogram represents the number of tokens required to encode 100 million polymers in SELFIES format, used for pre-training the language model.
        }
	\label{token_count} 
\end{figure}

\begin{figure}
	\centering
	\includegraphics[width=0.7\textwidth]{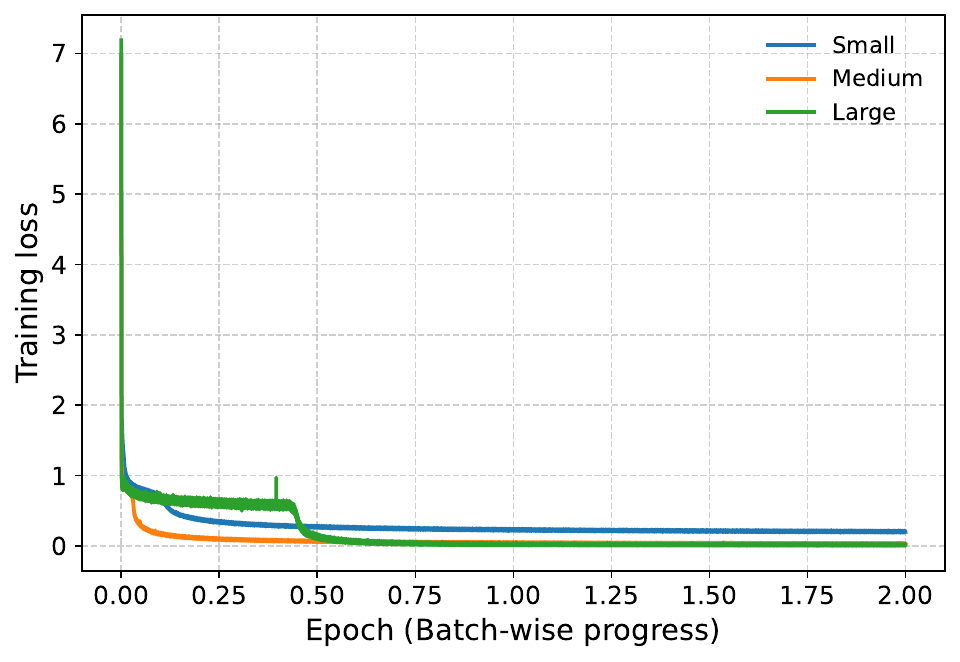}
        \caption{
        \textbf{Training loss per batch as a function of epoch progress for T5-based models of different sizes.}
         The x-axis represents fractional progress through training epochs, and the y-axis shows the mean cross-entropy loss computed over each batch. Results are shown for small, medium, and large model variants to compare training dynamics across model scales.
        }
	\label{train_loss} 
\end{figure}

\begin{figure}
	\centering
	\includegraphics[width=0.95\textwidth]{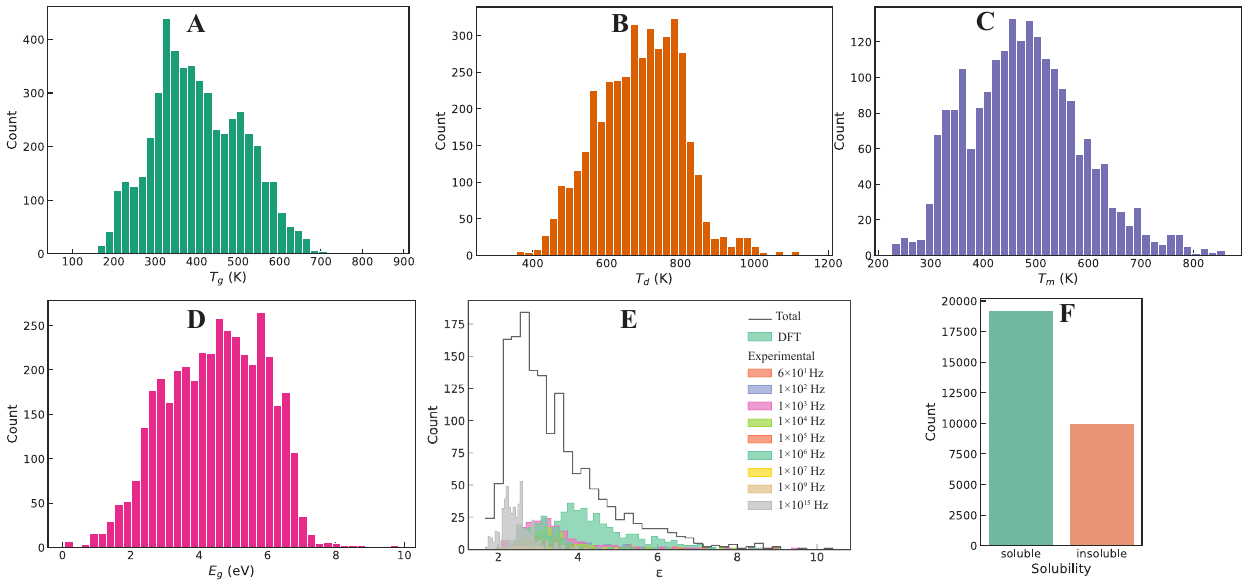}
        \caption{
        \textbf{Property distributions of polymer datasets.}
        Histogram plots showing the distribution of (\textbf{A}) glass transition temperature ($T_{\rm g}$, K), (\textbf{B}) decomposition temperature ($T_{\rm d}$, K), (\textbf{C}) melting temperature ($T_{\rm m}$, K), (\textbf{D}) electronic bandgap ($E_{\rm g}$, eV), and (\textbf{E}) dielectric constant ($\varepsilon$). Panel (\textbf{F}) presents the counts of soluble and insoluble cases in the polymer solubility dataset.
        }
	\label{prop_train} 
\end{figure}

\begin{figure}
	\centering
	\includegraphics[width=0.9\textwidth]{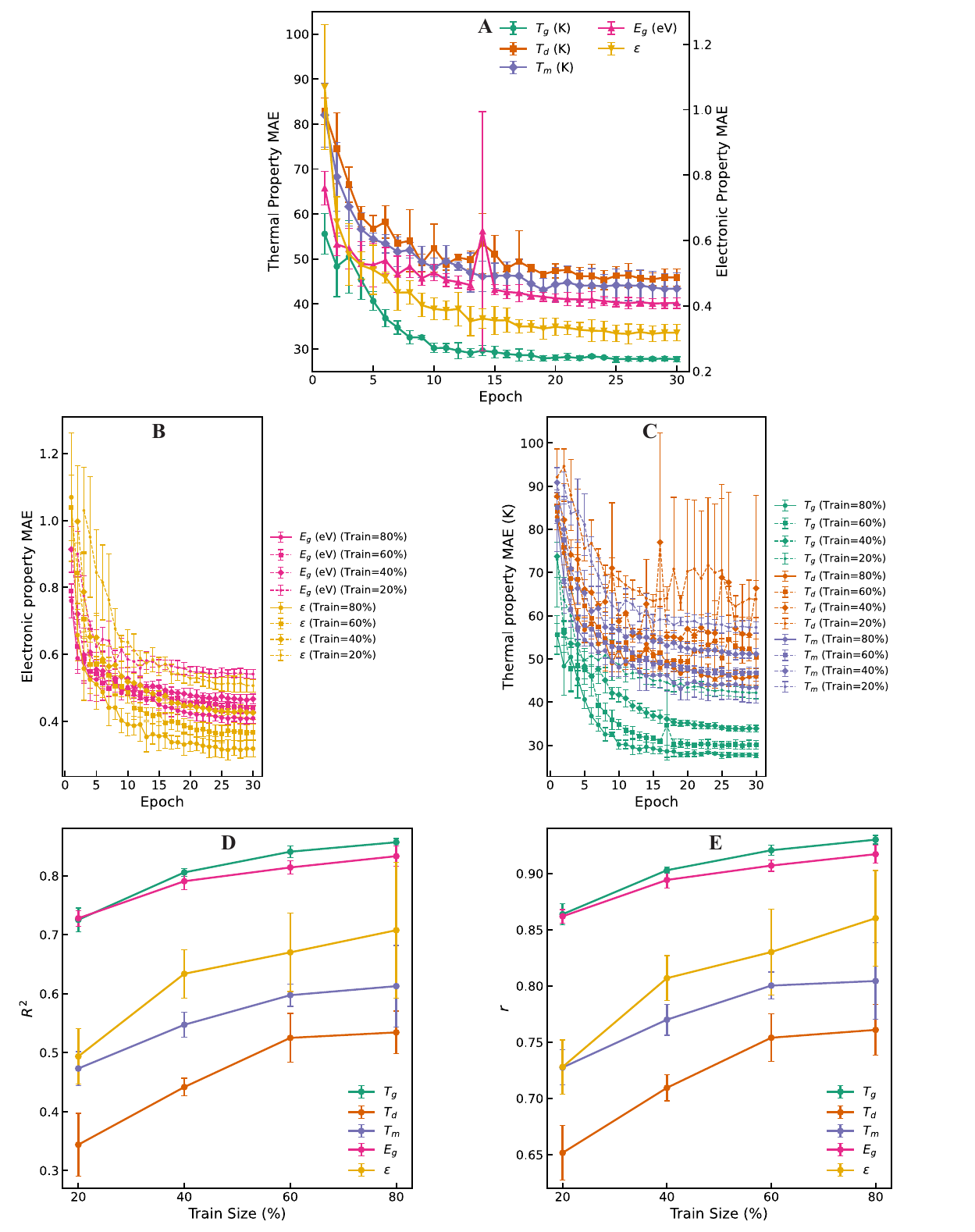}
        \caption{\textbf{Performance of fine-tuned \textsc{polyT5}-medium models for thermal and electronic property prediction.}
        (\textbf{A}) Mean absolute error (MAE) as a function of training epochs for an 80-20\% train-test split.
        (\textbf{B--C}) MAE across varying train-test splits, from 20-80\% to 80-20\%.
        (\textbf{D--E}) Learning curves showing the coefficient of determination ($R^2$) and Pearson correlation coefficient ($r$), respectively.
        For panels \textbf{A--E}, error bars represent the standard deviation across five different random data splits.
        }
	\label{figS4} 
\end{figure}

\begin{figure}
	\centering
	\includegraphics[width=0.9\textwidth]{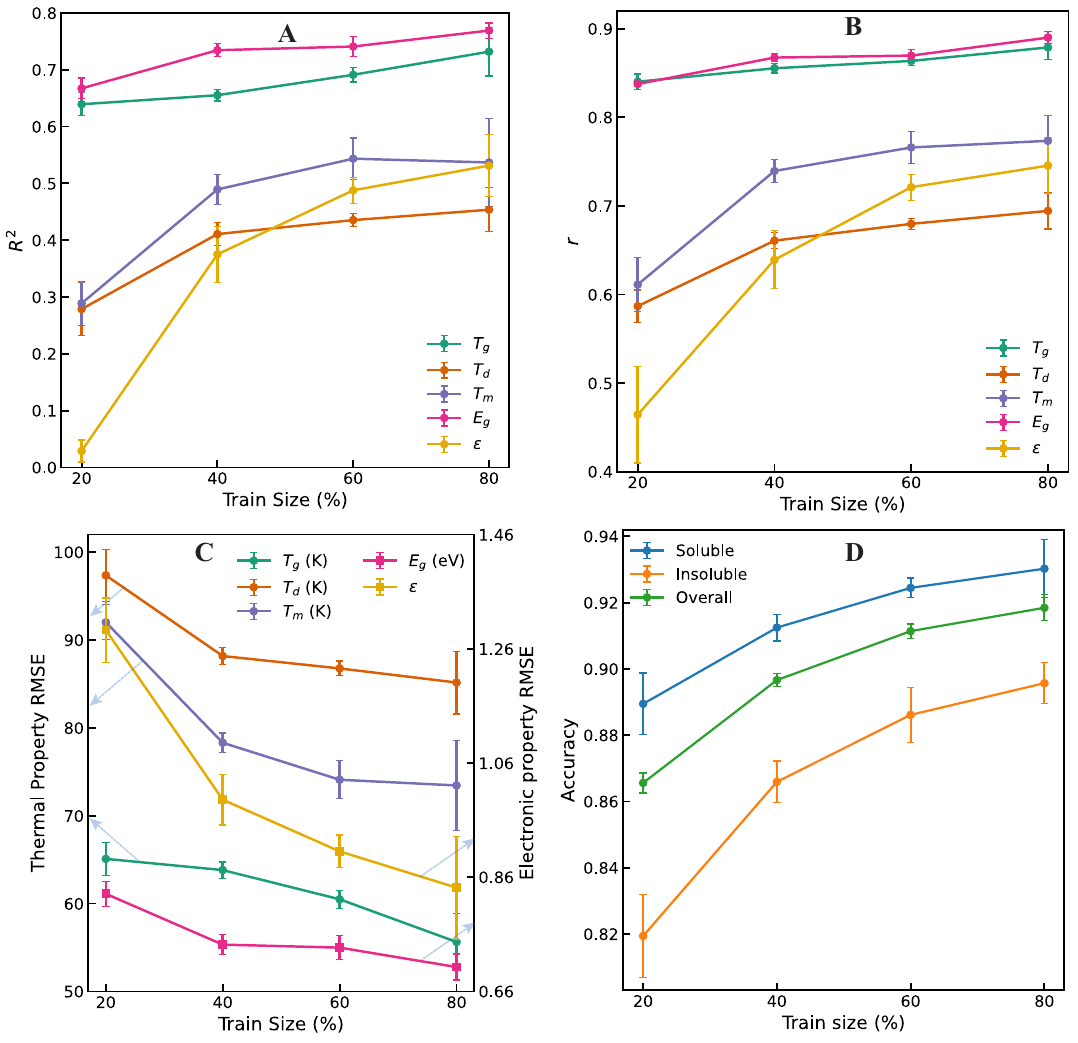}
        \caption{\textbf{Performance of fine-tuned \textsc{polyT5}-small models for thermal, electronic, and solubility property prediction.}
        (\textbf{A--C}) Learning curves showing the coefficient of determination ($R^2$), Pearson correlation coefficient ($r$), and root-mean-square error (RMSE), respectively, for thermal and electronic properties.
        (\textbf{D}) Learning curves for solubility prediction.
        Error bars indicate the standard deviation across five random data splits.
        }
	\label{figS5} 
\end{figure}

\begin{figure}
	\centering
	\includegraphics[width=0.9\textwidth]{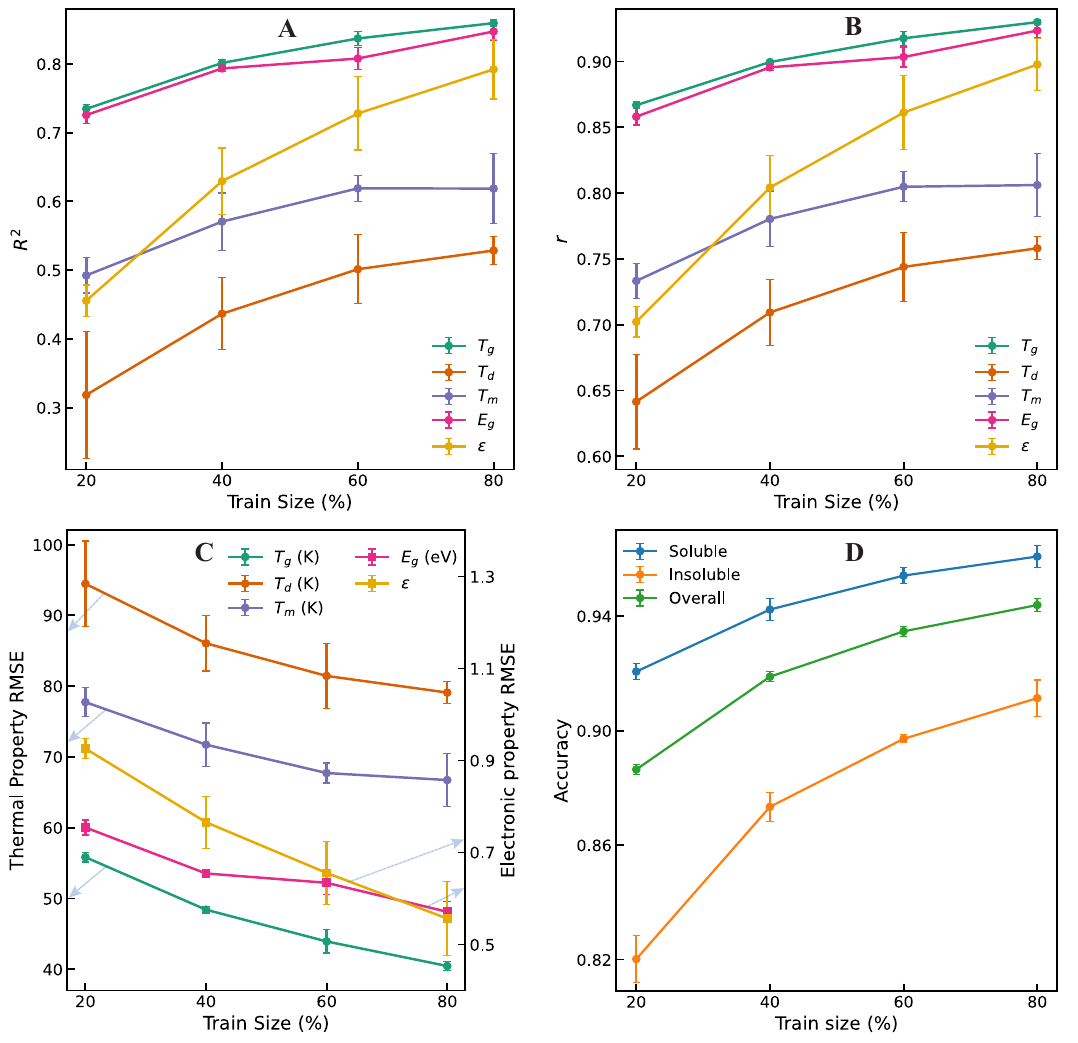}
        \caption{\textbf{Performance of fine-tuned \textsc{polyT5}-large models for thermal, electronic, and solubility property prediction.}
        (\textbf{A--C}) Learning curves showing the coefficient of determination ($R^2$), Pearson correlation coefficient ($r$), and root-mean-square error (RMSE), respectively, for thermal and electronic properties.
        (\textbf{D}) Learning curves for solubility prediction.
        Error bars indicate the standard deviation across five random data splits.
        }
	\label{figS6} 
\end{figure}

\begin{figure}
	\centering
	\includegraphics[width=0.85\textwidth]{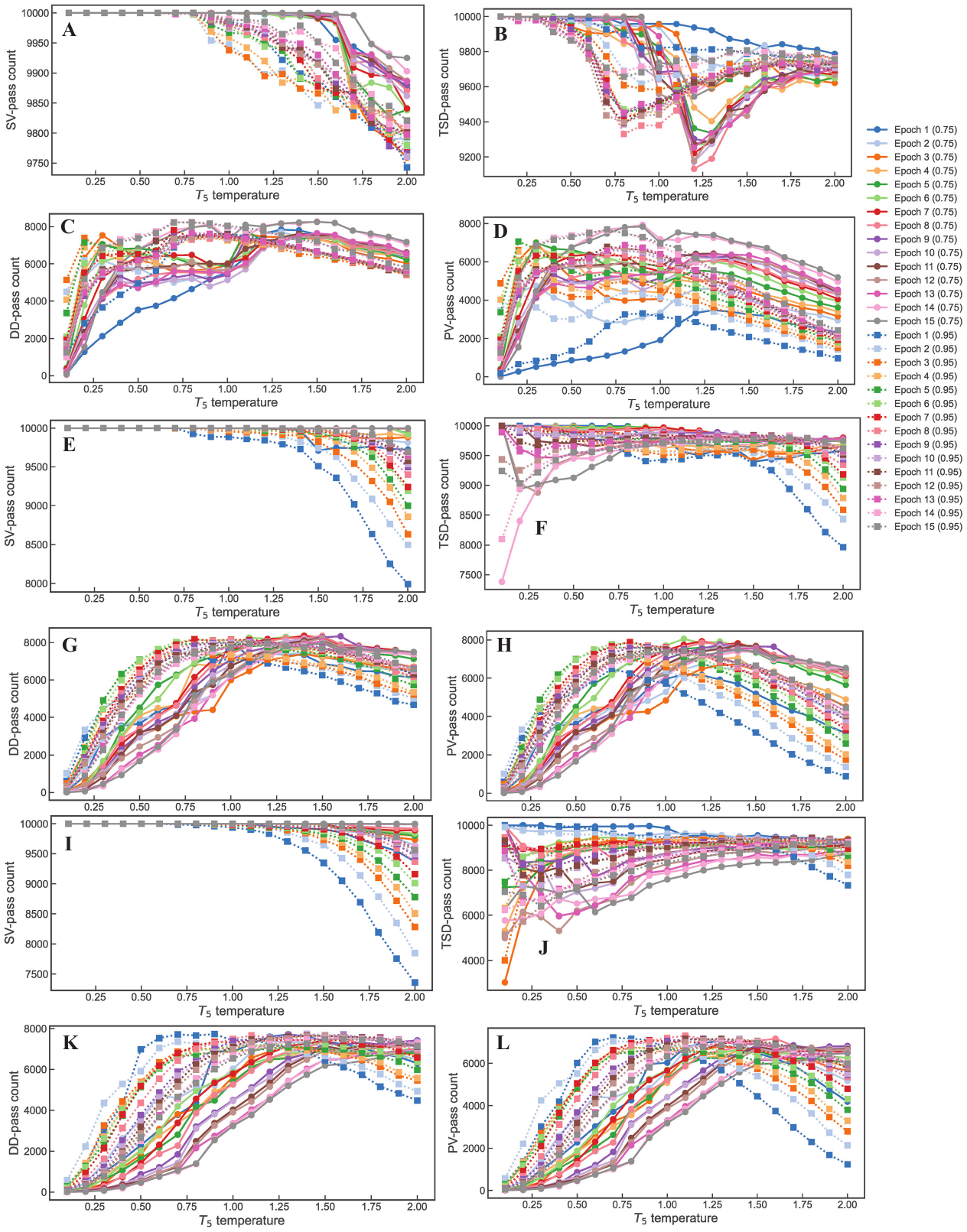}
        \caption{
        \textbf{Validity assessment of hypothetical polymers generated by \textsc{polyT5} across fine-tuning epochs, T5 temperatures, and sampling rates.}
        Candidates were filtered sequentially by SMILES validity (SV), training set deduplication (TSD), internal deduplication (DD), and polymer validity (PV) ensuring exactly two Astatine (At) atoms with valency one (PV~$\subset$ DD $\subset$ TSD $\subset$ ~SV).
        (\textbf{A--D}) show the number of candidates passing SV, TSD, DD, and PV checks, respectively, out of 10,000 generated for each case with \textsc{polyT5}-small; (\textbf{E--H}) and (\textbf{I--L}) correspond to \textsc{polyT5}-medium and \textsc{polyT5}-large.
        }
	\label{figS1} 
\end{figure}

\begin{figure}
	\centering
	\includegraphics[width=0.9\textwidth]{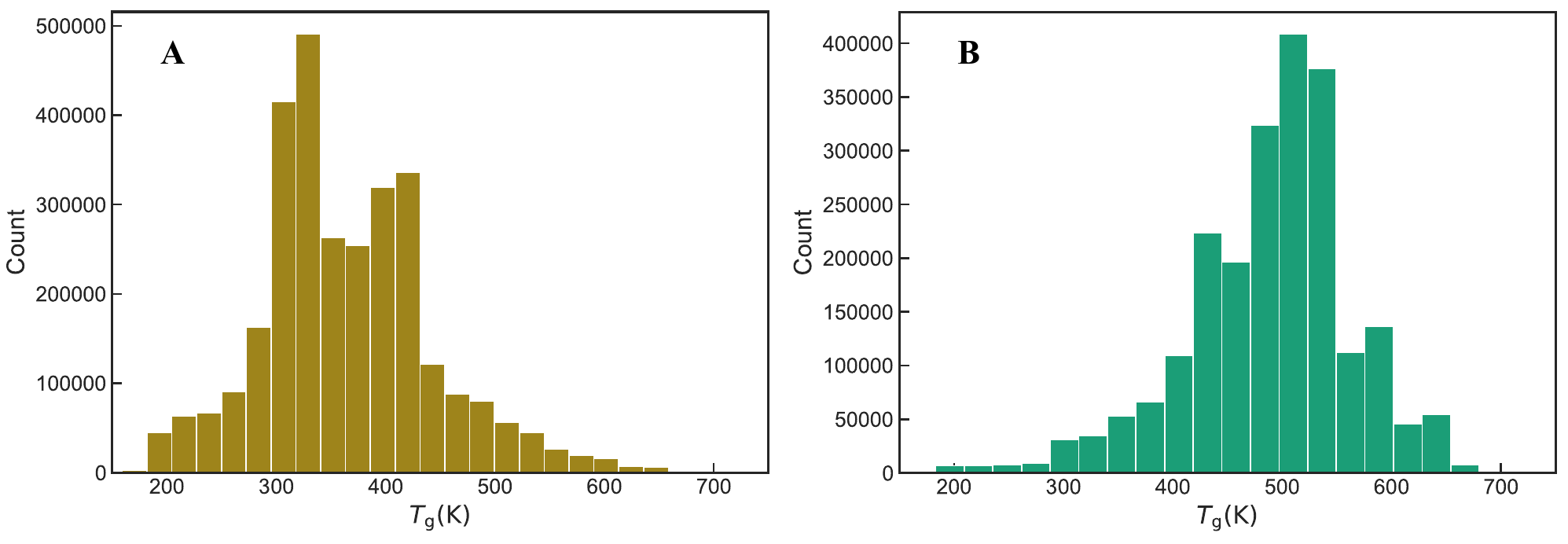}
        \caption{\textbf{Distribution of predicted $T_{\rm g}$ for hypothetical polymers generated by \textsc{polyT5}.} Hypothetical polymers conditioned on target temperatures of 300\,K (A) and 500\,K (B) show distinct $T_{\rm g}$ distributions, demonstrating the model’s ability to align generation with thermal property targets.}
	\label{fig300K} 
\end{figure}

\begin{figure}
	\centering
	\includegraphics[width=0.9\textwidth]{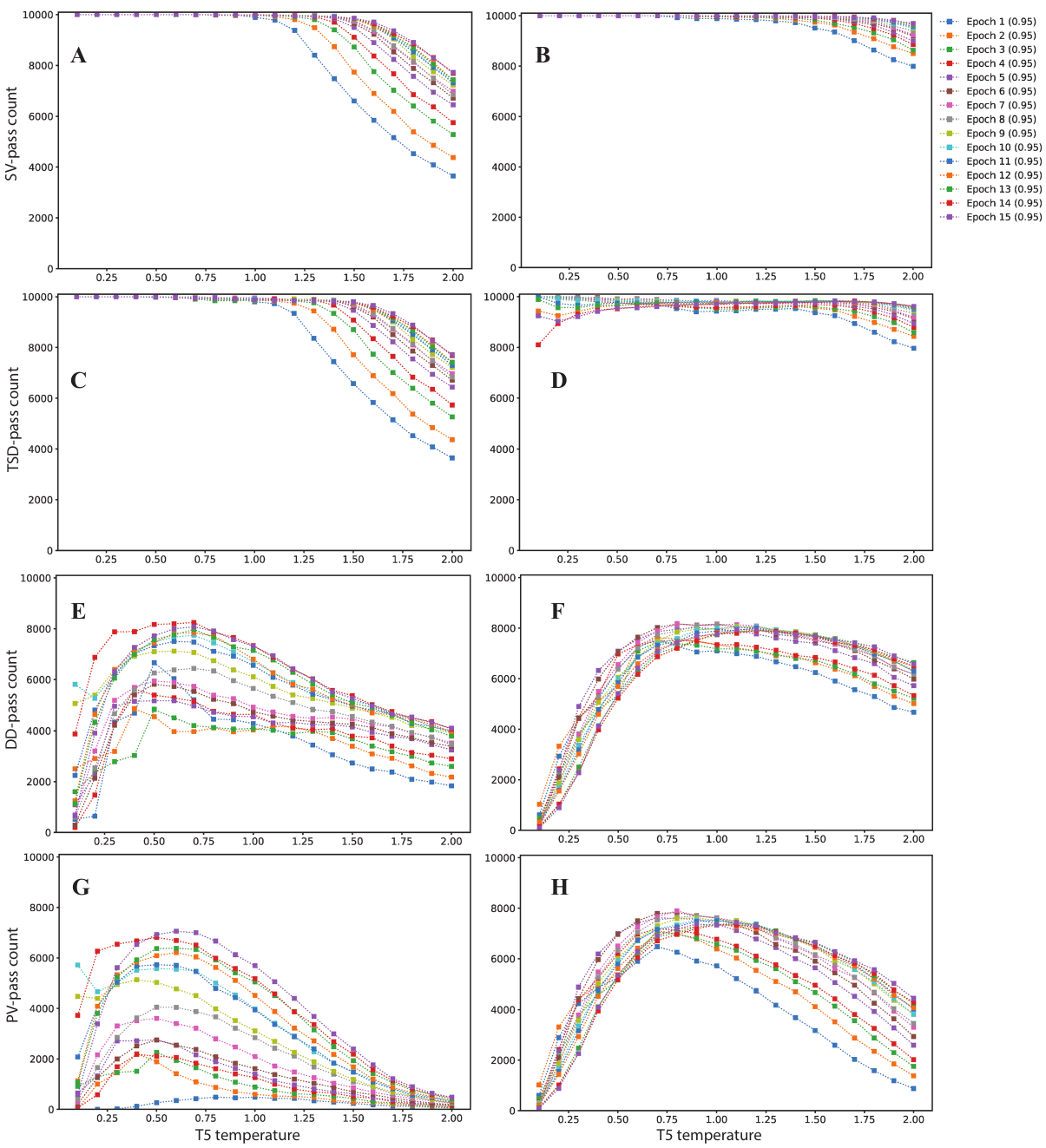}
        \caption{\textbf{Ablation study comparing pre-trained and non-pre-trained \textsc{polyT5}-medium models for generating hypothetical polymers targeting $T_{\rm g} = 500$\,K.} Validity of 10,000 generated candidates was assessed across fine-tuning epochs, temperature settings, and sampling rates. Candidates were sequentially filtered by SMILES validity (SV), training set deduplication (TSD), internal deduplication (DD), and polymer validity (PV), where PV $\subset$ DD $\subset$ TSD $\subset$ SV. Subfigures \textbf{A}, \textbf{C}, \textbf{E}, and \textbf{G} correspond to models without pre-training; \textbf{B}, \textbf{D}, \textbf{F}, and \textbf{H} show results with pre-training.}

	\label{abl_gen_part1} 
\end{figure}

\begin{figure}
	\centering
	\includegraphics[width=0.9\textwidth]{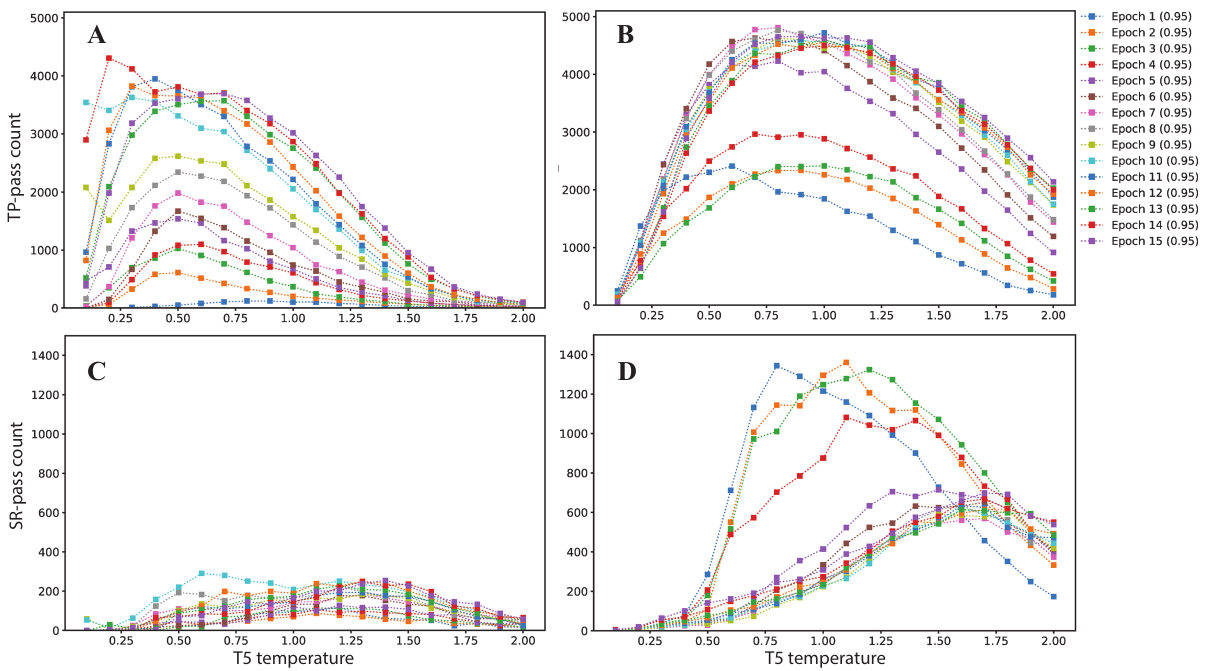}
        \caption{\textbf{Ablation study comparing pre-trained and non-pre-trained \textsc{polyT5}-medium models for generating polymers targeting $T_{\rm g} = 500$\,K, evaluated using property and reproducibility metrics.} Each model generated 10,000 hypothetical candidates. Two independent metrics were assessed: (1) TP — proportion of candidates with predicted $T_{\rm g}$ within 500\,$\pm$\,50\,K, and (2) SR — proportion passing the SELFIES reproducibility test. Subfigures \textbf{A} and \textbf{B} correspond to models without pre-training; \textbf{C} and \textbf{D} show results with pre-training.}

	\label{abl_gen_part2} 
\end{figure}

\begin{figure}
	\centering
	\includegraphics[width=0.5\textwidth]{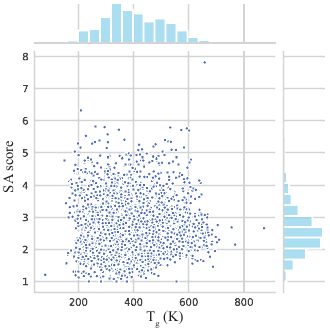}
        \caption{\textbf{SA score vs. $T_{\rm g}$ distribution for known polymers.} Distribution of synthetic accessibility (SA) scores for known polymers in the $T_{\rm g}$ dataset plotted against their respective $T_{\rm g}$ values, with accompanying marginal histograms showing the individual distributions of $T_{\rm g}$ and SA scores.} 
	\label{figS7} 
\end{figure}

\begin{figure}
	\centering
	\includegraphics[width=1.0\textwidth]{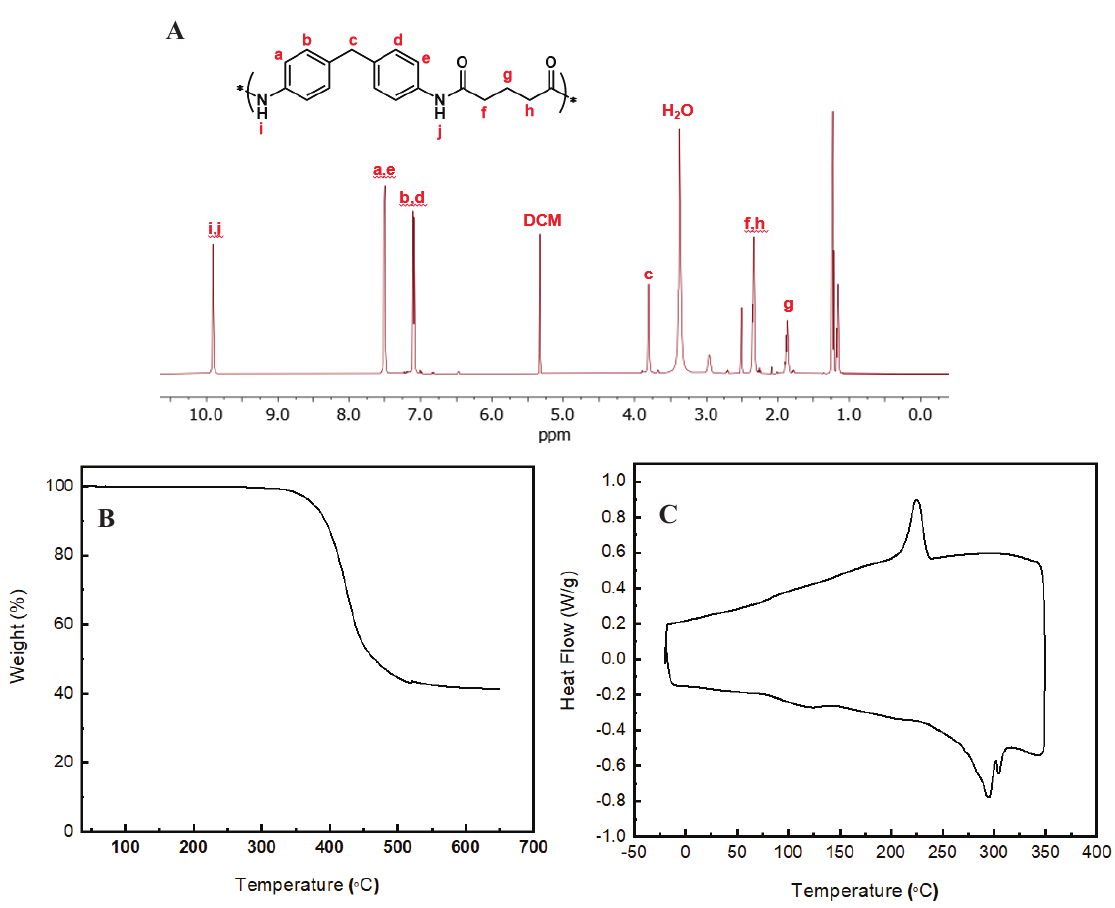}
        \caption{\textbf{Characterization of the synthesized polymer.} (\textbf{A}) $^1$H NMR spectra recorded in d$_{6}$-DMSO, showing characteristic proton signals confirming polymer structure. (\textbf{B}) Thermogravimetric analysis (TGA) curve demonstrating thermal stability and decomposition profile. (\textbf{C}) Differential scanning calorimetry (DSC) curve indicating thermal transitions such as glass transition and melting temperatures.} 
	\label{exp_results_SI} 
\end{figure}

\begin{figure} 
	\centering
	\includegraphics[width=0.98\textwidth]{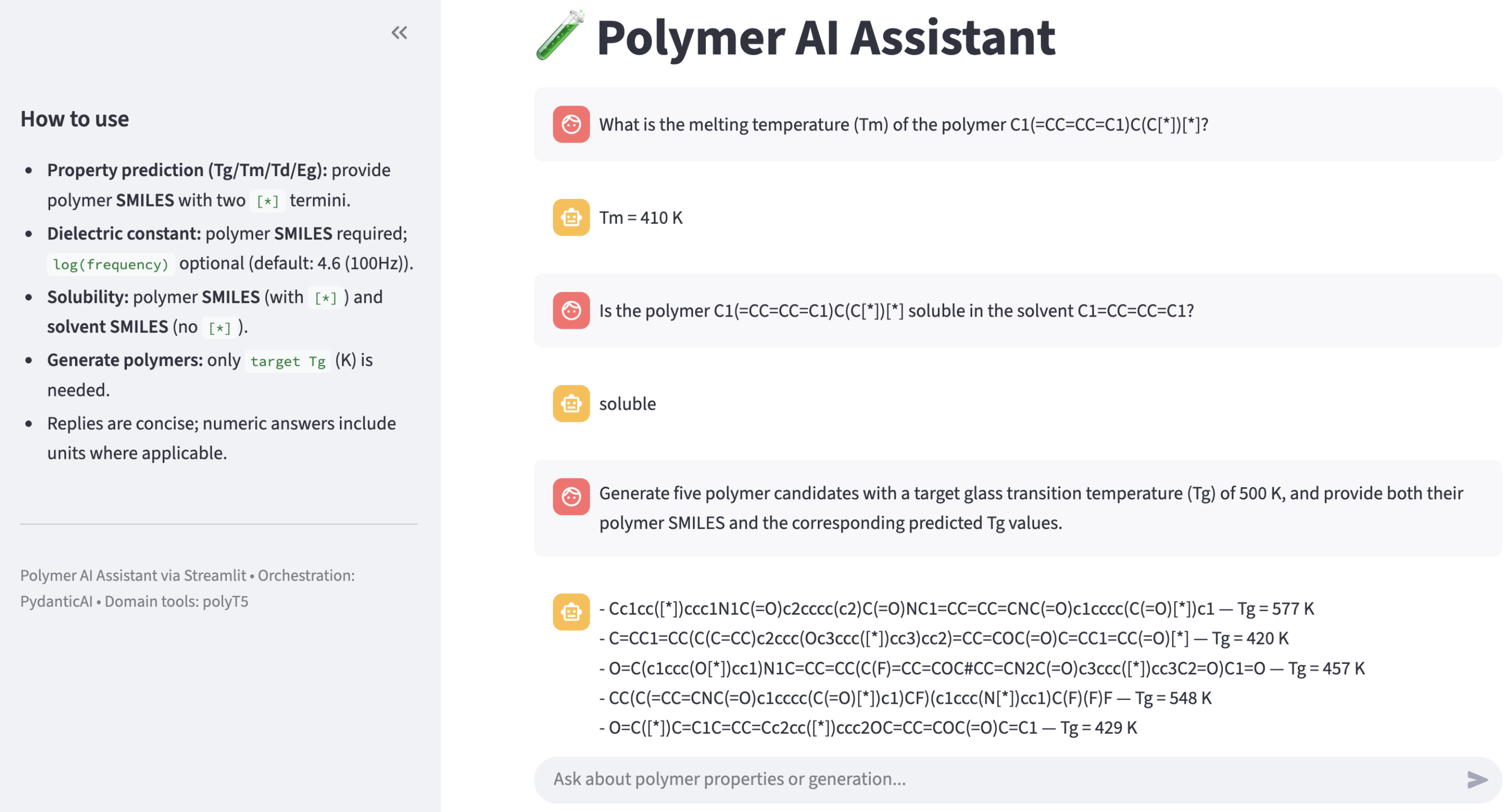} 
        \caption{
        \textbf{Snapshot of the agentic AI framework in action.}
        The user queries the chatbot in natural language and the system responds with predictions or generated candidates. This unified conversational interface integrates the general-purpose LLM controller with fine-tuned \textsc{polyT5} models, automatically handling input parsing, format conversion, and property prediction or generative design.
        }
	\label{chatbot}
\end{figure}

\clearpage
\begin{table} 
	\centering
	\caption{\textbf{Classification of the 100 million polymers used for pre-training \textsc{polyT5} (total counts = 100,145,883).} Polymers were categorized based on their elemental composition, degree of unsaturation, aromaticity, and the presence of functional groups. Classification was performed by considering the entire polymer structure, allowing a single polymer to belong to multiple categories.}
	\label{tabS1} 
	\begin{tabular}{lc|cc|cr} 
		\\
		\hline
		Elemental Composition & \% & Unsaturation/Aromaticity & \% & Functional Group & \%\\
		\hline
		Hydrocarbon & $<$0.1 & Any unsaturation & 99.8 & Ether & 90.7\\
		Non-hydrocarbon & 100.0 & Carbon unsaturation & 98.3 & Amide & 71.4 \\
		Oxygen & 99.0 & Aromatic ring & 97.9 & Amine & 65.1\\
            Nitrogen & 93.6 & Polycyclic Aromatic & 60.7 & Acetal & 30.3\\
            Sulfur & 41.1 &  &  & Thioether & 28.9\\
            Halogens & 34.0 &  &  & Ester & 23.1\\
            Phosphorus & 1.6 &  &  & Allyl & 21.4\\
            Silicon & 0.4 &  &  & Hydroxyl & 18.9\\
             &  &  &  &  Sulfoxide & 15.2 \\
             &  &  &  &  Urea & 11.1 \\
             &  &  &  &  Imide & 8.4 \\
             &  &  &  &  Nitrile & 5.7 \\
             &  &  &  &  Urethane & 5.3\\
             &  &  &  &  Carboxylic acid & 5.2\\
             &  &  &  &  Amidine & 4.1 \\
             &  &  &  &  Acrylates & 3.7 \\
             &  &  &  &  Hydrazides & 3.4 \\
             &  &  &  &  Alkyne & 2.2 \\
             &  &  &  &  Methacrylates & 2.0 \\
             &  &  &  &  Thiol & 1.5 \\
		\hline
	\end{tabular}
\end{table}

\begin{table} 
	\centering
	\caption{\textbf{Architectural configurations of the pre-trained T5 models.} The models differ in embedding dimension (d\_model), number of transformer layers (num\_layers), size of the feedforward network (d\_ff), number of attention heads (num\_heads), and dimensionality of key and value vectors (d\_kv). All models use a maximum sequence length (n\_positions) of 200 positions.}
	\label{tabS0} 
	\begin{tabular}{l|ccc} 
		\hline
		\textbf{Parameter} & \textbf{Small} & \textbf{Medium} & \textbf{Large} \\
		\hline
		d\_model & 128 & 256 & 512 \\
		num\_layers & 3 & 4 & 8 \\
		d\_ff & 512 & 1024 & 2048 \\
            num\_heads & 4 & 4 & 8\\
            d\_kv & 32 & 64 & 64 \\
            n\_positions & 200 & 200 & 200 \\
            \hline
            \textbf{Total parameters (million)} & 1.44 & 7.46 & 58.98\\
		\hline
	\end{tabular}
\end{table}

\begin{table} 
	\centering
	\caption{\textbf{Classification of polymers in glass-transition temperature ($T_{\rm g}$) dataset (total counts = 5,130) used for fine-tuning \textsc{polyT5} for hypothetical candidate generation.} Polymers were categorized based on their elemental composition, degree of unsaturation, aromaticity, and the presence of functional groups. Classification was performed by considering the entire polymer structure, allowing a single polymer to belong to multiple categories.}
	\label{tabS2} 
	\begin{tabular}{lc|cc|cr} 
		\\
		\hline
		Elemental Composition & \% & Unsaturation/Aromaticity & \% & Functional Group & \%\\
		\hline
		Hydrocarbon & 3.2 & Any unsaturation & 94.5 & Ether & 35.6\\
		Non-hydrocarbon & 96.8 & Carbon unsaturation & 78.8 & Amide & 34.7 \\
		Oxygen & 88.5 & Aromatic ring & 76.0 & Ester & 34.2 \\
            Nitrogen & 43.0 & Polycyclic Aromatic & 46.8 & Imide & 13.9\\
            Halogens & 14.4 &  &  & Thioether & 8.9\\
            Sulfur & 10.8 &  &  & Sulfoxide & 6.9\\
            Silicon & 4.1 &  &  & Allyl & 5.4\\
            Phosphorus & 3.4 &  &  & Amine & 4.9\\
             &  &  &  &  Hydroxyl &  4.9\\
             &  &  &  &  Nitrile & 3.3 \\
             &  &  &  &  Urethane &  3.2\\
             &  &  &  &  Urea &  1.0 \\
             &  &  &  &  Epoxide & 1.0 \\
             &  &  &  &  Phosphate & 0.8 \\
             &  &  &  &  Acetal &  0.8 \\
             &  &  &  &  Anhydride &  0.8 \\
             &  &  &  &  Acrylates &  0.7 \\
             &  &  &  &  Carboxylic acid &  0.7\\
             &  &  &  &  Hydrazides & 0.5 \\
             &  &  &  &  Alkyne &  0.4 \\
		\hline
	\end{tabular}
\end{table}

\begin{table} 
	\centering
	\caption{\textbf{Classification of hypothetical polymers generated using \textsc{polyT5} with targeted glass-transition temperature ($T_{\rm g}$) of 500~K (total counts = 6,171,066).} Polymers were categorized based on their elemental composition, degree of unsaturation, aromaticity, and the presence of functional groups. Classification was performed by considering the entire polymer structure, allowing a single polymer to belong to multiple categories.}
	\label{tabS4} 
	\begin{tabular}{lc|cc|cr} 
		\\
		\hline
		Elemental Composition & \% & Unsaturation/Aromaticity & \% & Functional Group & \%\\
		\hline
		Hydrocarbon & 0.7 & Any unsaturation & 98.9 & Allyl & 84.3\\
		Non-hydrocarbon & 99.3 & Carbon unsaturation & 94.6 & Ether & 66.1 \\
		Oxygen & 96.6 & Aromatic ring & 85.8 & Amide & 56.0 \\
            Nitrogen & 67.1 & Polycyclic Aromatic & 76.5 & Ester & 24.2\\
            Halogens & 11.8 &  &  & Imide & 14.0 \\
            Sulfur & 11.6 &  &  & Amine & 12.8\\
            Phosphorus & 2.4 &  &  & Aldehyde & 10.6 \\
            Silicon & 2.0 &  &  & Acrylates & 9.5\\
             &  &  &  &  Thioether & 9.2 \\
             &  &  &  &  Sulfoxide & 7.5 \\
             &  &  &  &  Hydroxyl & 6.1 \\
             &  &  &  &  Vinylsulfone & 5.4  \\
             &  &  &  &  Urethane & 3.5 \\
             &  &  &  &  Methacrylates & 3.4 \\
             &  &  &  &  Urea & 2.6  \\
             &  &  &  &  Nitrile & 2.2 \\
             &  &  &  &  Amidine &  1.6 \\
             &  &  &  &  Acetal & 1.2 \\
             &  &  &  &  Hydrazides & 1.0 \\
             &  &  &  &  Alkyne & 0.7  \\
		\hline
	\end{tabular}
\end{table}

\begin{table} 
	\centering
        \caption{\textbf{Ablation study comparing model performance with and without pre-training across various properties.} In the non-pre-trained setting, model weights were randomized while keeping the architecture and training protocol identical, including fine-tuning for 30 epochs. Metrics reported include RMSE, $R^2$, and Pearson’s $r$ for regression tasks, and soluble accuracy, insoluble accuracy, and overall accuracy for classification. Values in parentheses represent standard deviations computed over five different data splits.}
	\label{tabS5} 
        \begin{tabular}{lccc}
        \hline
        \textbf{Property} & \textbf{RMSE} & \textbf{R\textsuperscript{2}} & \textbf{r} \\
        \hline
        $T_{\rm g}$ (pre\_train) & 40.82 (1.33) & 0.86 (0.00) & 0.93 (0.01) \\
        $T_{\rm g}$              & 89.35 (4.65) & 0.31 (0.05) & 0.65 (0.01) \\
        \hline
        $T_{\rm d}$ (pre\_train) & 78.59 (3.56) & 0.53 (0.03) & 0.76 (0.02) \\
        $T_{\rm d}$              & 113.59 (8.96) & 0.02 (0.16) & 0.39 (0.07) \\
        \hline
        $T_{\rm m}$ (pre\_train) & 67.07 (5.18) & 0.61 (0.07) & 0.80 (0.03) \\
        $T_{\rm m}$              & 175.81 (61.02) & -1.96 (1.77) & 0.25 (0.21) \\
        \hline
        $E_{\rm g}$ (pre\_train) & 0.60 (0.03) & 0.83 (0.02) & 0.92 (0.01) \\
        $E_{\rm g}$              & 0.92 (0.05) & 0.60 (0.04) & 0.79 (0.02) \\
        \hline
        $\varepsilon$ (pre\_train) & 0.65 (0.10) & 0.71 (0.11) & 0.86 (0.04) \\
        $\varepsilon$              & 0.88 (0.09) & 0.49 (0.05) & 0.73 (0.03) \\
        \hline
        \hline
         & \textbf{Soluble Accuracy} & \textbf{Insoluble Accuracy} & \textbf{Accuracy} \\
        \hline
        Solubility (pre\_train) & 0.96 (0.00) & 0.92 (0.01) & 0.94 (0.00) \\
        Solubility              & 0.98 (0.02) & 0.04 (0.04) & 0.66 (0.00) \\
        \hline
        \end{tabular}
\end{table}


\clearpage 





\end{document}